\documentclass[amssymb,amsmath,prd,twocolumn,superscriptaddress,nofootinbib]{revtex4-1}

\usepackage{graphicx}
\usepackage{bm}
\usepackage{amssymb,amsmath}
\usepackage{mathrsfs}
\usepackage{latexsym}
\usepackage{color}
\usepackage[normalem]{ulem}
\usepackage{dcolumn}
\usepackage[colorlinks=true,citecolor=blue,urlcolor=blue]{hyperref}
\usepackage[usenames,dvipsnames]{xcolor}
\usepackage{xspace}

\allowdisplaybreaks

\newcommand{\CCA}{\affiliation{Center for Computational Astrophysics, Flatiron Institute, 162 5th Ave, New York, NY 10010, USA}}
\newcommand{\LIGOlabMIT}{\affiliation{LIGO Laboratory, Massachusetts Institute of Technology, 185 Albany St, Cambridge, MA 02139, USA}}
\newcommand{\MKI}{\affiliation{MIT-Kavli Institute for Astrophysics and Space Research, 77 Massachusetts Ave, Cambridge, MA 02139, USA}}
\newcommand{\MSFC}{\affiliation{NASA Marshall Space Flight Center, Huntsville, AL 35812, USA}}
\newcommand{\SBU}{\affiliation{Department of Physics and Astronomy, Stony Brook University, Stony Brook NY 11794, USA}}
\newcommand{\XGI}{\affiliation{eXtreme Gravity Institute, Department of Physics, Montana State University, Bozeman, Montana 59717, USA}}
\newcommand{\Nice}{\affiliation{Astrophysique Relativiste, Th\'eories, Exp\'eriences, M\'etrologie, Instrumentation, Signaux (ARTEMIS), Bd de l'Observatoire, B.P. 4229, 06304 Nice CEDEX 4, France}}
\newcommand{\MelbourneOzGrav}{\affiliation{OzGrav, University of Melbourne, Parkville, Victoria 3010, Australia}}
\newcommand{\GT}{\affiliation{Center for Relativistic Astrophysics and School of Physics, Georgia Institute of Technology, Atlanta, GA 30332, USA}}

\definecolor{kcmagenta}{rgb}{0.54, 0.17, 0.88}
\definecolor{chorange}{rgb}{0.851, 0.372, 0.007}
\definecolor{tlteal}{rgb}{0,.55,.55}

\newcommand{\BayesLine}{{\tt BayesLine}\xspace}
\newcommand{\BayesWave}{{\tt BayesWave}\xspace}
\newcommand{\LALInference}{{\tt LALInference}\xspace}
\newcommand{\SNR}{S/N\xspace}
\newcommand{\normal}{{\cal{N}}(0,1)}

\begin{document}

\title{Noise spectral estimation methods and their impact on gravitational wave measurement of compact binary mergers}

\author{Katerina Chatziioannou} \CCA
\author{Carl-Johan Haster} \LIGOlabMIT \MKI
\author{Tyson B. Littenberg}\MSFC
\author{Will M. Farr} \SBU \CCA
\author{Sudarshan Ghonge} \GT
\author{Margaret Millhouse} \MelbourneOzGrav
\author{James A. Clark} \GT 
\author{Neil Cornish} \XGI \Nice

\date{\today}

\begin{abstract}
Estimating the parameters of gravitational wave signals detected by ground-based detectors requires an understanding of the properties of the detectors' noise.
In particular, the most commonly used likelihood function for gravitational wave data analysis assumes that the noise is Gaussian, stationary, and of known frequency-dependent variance.
The variance of the colored Gaussian noise is used as a whitening filter on the data before computation of the likelihood function.
In practice the noise variance is not known and it evolves over timescales of dozens of seconds to minutes.
We study two methods for estimating this whitening filter for ground-based gravitational wave detectors with the goal of performing parameter estimation studies.
The first method uses large amounts of data separated from the specific segment we wish to analyze and computes the power spectral density of the noise through the mean-median Welch method.
The second method uses the same data segment as the parameter estimation analysis, which potentially includes a gravitational wave signal, and obtains the whitening filter through a fit of the power spectrum of the data in terms of a sum of splines and Lorentzians.
We compare these two methods and conclude that the latter is a more effective spectral estimation method as it is quantitatively consistent with the statistics of the data used for gravitational wave parameter estimation while the former is not. We demonstrate the effect of the two methods by finding quantitative differences in the inferences made about the physical properties of simulated gravitational wave sources added to LIGO-Virgo data.
\end{abstract}

\maketitle

\section{Introduction}
\label{intro}

Analysis of data containing compact binary coalescence (CBC) signals from ground-based gravitational wave (GW) detectors~\cite{TheLIGOScientific:2014jea,TheVirgo:2014hva} relies on accurate models not only of the expected signal waveforms, but also of the detector noise~\cite{Creighton:2011zz}.
While the field of waveform modeling has received extensive attention over the last decades~\cite{Blanchet:2014zz}, modeling the detector noise has been historically less mainstream.
Despite this, the properties of the detector noise have been at the forefront of investigations around GW detections, including the first binary neutron star (BNS) observation, GW170817, as the signal overlapped with a major noise excursion in one of the detectors~\cite{TheLIGOScientific:2017qsa}.

Traditional template-based analyses of gravitational wave signals depend on noise weighted inner products of waveform models and data which, in turn, depend on three assumptions about the random noise processes in the data:
\begin{enumerate}
\item The noise is Gaussian, completely characterized by a mean vector and a covariance matrix. 
\item The noise is stationary, i.e. the mean and covariance do not change in time. In the frequency domain, the covariance matrix is diagonal, and completely characterized by the noise variance.
\item The frequency-dependent variance of the noise is known.
\end{enumerate}
Of particular relevance to this work, all three of these assumptions are explicitly invoked when evaluating the likelihood function in the \LALInference parameter estimation (PE) pipeline used by LIGO-Virgo~\cite{Veitch:2014wba,lalsuite}.

However, all three assumptions are invalid to one extent or another.
Most obviously, the variance of the noise is not known \emph{a priori}, and needs to be estimated from the available data, possibly in a way that incorporates the uncertainty in that estimation \cite{Rover:2008yp}.
Secondly, the noise process is time evolving~\cite{Aasi:2013jjl}, though for transient sources it is a reasonable approximation that the stationary timescales of the noise are long compared to the stride of data containing the signals~\cite{TheLIGOScientific:2016zmo,Cahillane:2017vkb}.
Finally, non-Gaussian noise excursions, or ``glitches'', are common in the detectors~\cite{TheLIGOScientific:2016zmo,Zevin:2016qwy} and can potentially occur in data also containing signals.
One such noise excursion, an overflow glitch, overlapped with the BNS signal GW170817, explicitly breaking the Gaussianity assumption necessary for PE from pipelines available at the time.
In that case, the glitch was coherently fit and regressed from the data, leaving a Gaussian residual and enabling robust PE~\cite{TheLIGOScientific:2017qsa,Pankow:2018qpo}.

In this paper we revisit the assumptions about the noise variance being known and stationary, and compare
methods to estimate it as input to the LIGO-Virgo pipeline used for PE of CBC signals.
Our paper is framed as a comparative study between procedures that are in use for LIGO-Virgo parameter estimation, not as a complete survey and analysis of spectral estimation methods. We are therefore ignoring the broader landscape of spectral characterization algorithms, of which there are too many to enumerate here. To that end we focus on two approaches for estimating the noise variance.

The first method uses ``off-source'' data -- data near in time, but not containing, the detected signal -- to estimate the power spectral density (PSD) of the noise using a periodogram-based approach.
The off-source data are subdivided into segments equal to the duration $T$ of the data to be analyzed, and the PSD is estimated by averaging the power spectrum of the data from each segment, being careful to avoid biases due to large outliers or data windowing effects.
Details of the approach, referred to as the mean-median method, and its implementation in the \LALInference pipeline are described in Ref.~\cite{Veitch:2014wba}.

The second method uses only ``on-source'' data containing the signal and infers the frequency-dependent noise variance with a parametrized model.
The model for the noise variance is a two-component phenomenological fit.
The broadband noise is fit using a cubic spline where the number and location of control points for the spline are free parameters. 
This fit can be thought of as a Gaussian process regression for the smooth component of the PSD \cite{Kimeldorf:1970}, with each instance of the proposed spline model considered as a fair draw from the Gaussian process generating the noise.
Narrow-band features in the noise are fit using a linear combination of Lorentzians parametrized by their central frequency, amplitude, and line width.
Similar to the spline model, the number of Lorentzians in the fit is a free parameter, and models are explored using a transdimensional Markov Chain Monte Carlo (MCMC) algorithm, \BayesLine~\cite{Littenberg:2014oda}.

In this paper we expand upon the exploratory studies in~\cite{Littenberg:2014oda} and compare the two noise estimation
methods by performing a number of quantitative checks on a more extensive set of PE analyses.
In an effort to emulate the challenges faced by realistic PE for advanced ground-based detectors
we use real publicly available data from the two LIGO~\cite{TheLIGOScientific:2014jea} detectors and restrict to methods
as deployed in analysis of real signals. In particular, we simulate CBC signals in different mass regimes, inject them into the data,
and analyze them using LIGO-Virgo PE software~\cite{lalsuite} replicating the analysis procedures used in Ref.~\cite{LIGOScientific:2018mvr}.

We perform ``posterior predictive checks'' by testing whether the data conditioned by our estimates of the noise variance are consistent with the  underlying assumptions about the noise set out above.
We find that the on-source parametrized fits typically outperform the off-source estimation method in these tests.
We then show the effect of the noise variance estimation method on the inferred parameters of the simulated systems, demonstrating quantitative differences.
We conclude that, of the two methods tested here, the on-source method yields estimates of the noise that are more faithful to the foundational assumptions upon which current PE methods are built, and therefore is the preferred method for noise characterization.

We also perform exploratory checks on how PE results are impacted when the noise deviates from the stationarity assumption. We inject
CBC sources into LIGO data and analyze them with``on-source" whitening filters computed from data increasingly away from the injections.
We find that as the separation in time between the signal and the data used to compute the whitening filter increases, so do the differences in PE results.
We attribute these differences to noise nonstationarity and advise against using longer-than-necessary data segments in traditional PE analyses.

The rest of the paper presents the details of our study.
In Sec.~\ref{PSD} we describe the off-source and the on-source way of computing the noise variance as well as the simulated signals we analyze.
In Sec.~\ref{tests} we describe various tests we perform on the computed noise variances.
In Sec.~\ref{PE} we discuss the effect of the noise variance on PE from the CBC sources we simulate.
In Sec.~\ref{stationarity} we explore the effect of noise nonstationarity on PE results.
Finally, in Sec.~\ref{conclusions} we conclude.

\section{The noise variance}
\label{PSD}

The variance of noise in GW detectors is one of the ingredients necessary for computing the likelihood function of the data given a signal model.
In this section we describe the role of the noise variance in GW PE and the common ways of computing it.
We also discuss the CBC signals we simulate to test the properties of the methods for computing the noise variance.

\subsection{The role of the noise variance}

Our modeling assumption is that the data collected by ground-based GW detectors can be expressed as
\begin{equation}
d=h+n
\end{equation}
where $d$ is the data, $h$ is a GW signal that is coherent across the observatory network, and $n$ is the random noise independent in each detector.
Assuming an accurate model for the GW signal $h'$, the residual $r\equiv d-h'$ should have the same statistical properties as the detector noise.
Then the likelihood function $ {\cal{L}}(d|h')$ in a single detector, i.e. the probability (density) of measuring the data $d$ under the assumption that the true signal is $h'$ is the probability (density) of drawing $r$ from the noise distribution.
For Gaussian noise this reduces to
\begin{equation}
\mathrm{ln} {\cal{L}}(d|h') = -\frac{1}{2} r_i C_{ij}^{-1} r_j+{\rm const}\label{Lgaussian},
\end{equation}
where Einstein summation is assumed, the subscripts denote specific time or frequency bins, $C_{ij}\equiv\langle n_i n_j\rangle$ is the noise covariance matrix, we assume that the noise process is zero-mean, and the constant depends only on the covariance matrix and not on the data (or residual).
Angle brackets denote an average over noise realization.

If we further assume that the detector noise is stationary, i.e. its properties do not change on the timescales of interest, then the noise covariance matrix reduces to a diagonal matrix in the frequency domain
\begin{equation}
C_{ij}\equiv\langle \tilde{n}_i \tilde{n}_j\rangle = \frac{T}{2}S_n(f_i)\delta_{ij}\label{Lstationary},
\end{equation}
where no summation is assumed, $\delta_{ij}$ is the Kronecker delta function, $T$ is the duration of the analysis segment, and the overhead tilde marks frequency-domain quantities (e.g., see Appendix D in Ref \cite{Romano:2017}).
Now the (natural logarithm of the) likelihood function further reduces to~\cite{Veitch:2014wba}
\begin{equation}\label{Lfinal}
\mathrm{ln} {\cal{L}}(d|h')= -2 \sum_i^{N/2}\frac{\tilde{r}_i \tilde{r}_i^*}{ T S_n(f_i)} +{\rm const},
\end{equation}
where a star denotes the complex conjugate, $i$ counts the frequency bins, and $N$ is the number of time samples, equal to the sampling rate times $T$.
For a multiobservatory network with independent noise, the joint likelihood is the product of the individual likelihood functions for each detector's data.
This expression is the well-known likelihood function used for GW PE~\cite{Veitch:2014wba}
\footnote{The likelihood function is defined slightly differently in \BayesLine, specifically the factor of $T/2$ is absorbed in the definition of $S_n(f)$.
The end product for the numerical values of $\mathrm{ln} {\cal{L}}(d|h')$ is the same.}.

The function $S_n(f)$ is usually referred to as the PSD of the noise but, for our purposes here, it is simply thought of as a whitening filter, by which we divide the data (or residuals) with the expectation that the result will be consistent with a collection of fair draws from a zero-mean unit-variance Gaussian distribution, $\normal$.

The above introductory discussion to the derivation of the likelihood function highlights the importance of the three assumptions about the noise in the final expression:
\begin{itemize}
\item The Gaussian nature of the noise is invoked when requiring that the residuals are distributed according to a normal distribution, Eq.~\eqref{Lgaussian}.
\item The stationarity of the noise is required to express $C_{ij}$ as a diagonal matrix, Eq.~\eqref{Lstationary}, reducing the number of operations from $\mathcal{O}(N^2)$ to $\mathcal{O}(N)$ when evaluating the summation in Eq.~\eqref{Lgaussian}.
\item Finally, in order to compute the likelihood function we need to know $S_n(f)$.
\end{itemize}

The main focus of this paper is the whitening filter $S_n(f)$ and how to realistically compute it in the context of LIGO-Virgo PE of GW transients.
Given that the detector properties change with time, $S_n(f)$ needs to be computed for each detector and for each signal separately from the available data.
Before we turn into describing the two main ways to compute $S_n(f)$ used in LIGO-Virgo PE studies to date, we discuss some general considerations of the $S_n(f)$ calculation.

We separate the detector data in two pieces.
The first, denoted $d_s$, are the data that contain the signal and the assumed noise, and enter the numerator of the likelihood function.
The second, denoted $d_n$ are the data we will use to estimate $S_n(f)$, and hence enter the denominator of the likelihood function.
The stationarity assumption requires that neither segment of data is too long, otherwise the properties of the noise could change nontrivially.
For this reason, it is customary to choose $d_s$ to be as short as possible.
Specifically, $d_s$ is chosen to be the smallest power-of-2 integer number of seconds that include the entire duration of time that the signal spends in the measurement band of the detectors.

For example, analyses of high mass binary black holes (BBHs) such as GW150914 use $T=4$s~\cite{TheLIGOScientific:2016wfe}; for lower mass BBHs such as GW151226, $T=8$s~\cite{PhysRevLett.116.241103} is appropriate; and BNS signals such as GW170817 require $T=128$s, which contain the entire signal for a BNS with a chirp mass of ${\cal{M}}\sim 1.1975M_{\odot}$ (in the detector frame) from 23Hz to 2048Hz~\cite{Abbott:2018wiz}.

We then need to compute $S_n(f)$ from $d_n$.
We again require that $d_n$ is short enough such that the detector noise remains stationary.
We also require that $d_n$ is computed with the same spectral resolution $df$ as $d_s$.
This ensures that the spectral lines are resolved to the same accuracy in the numerator and the denominator of the likelihood, and hence do not affect PE considerably.

Below we describe the two main ways to compute $S_n(f)$.
One method uses a segment of data near in time to, but not including, the data that contain the GW signal itself, which we refer to as the \emph{off-source} method.
This approach assumes that the data collected adjacent to the detection are a good proxy for the noise behavior in the data to be analyzed.
This is analogous to using flat-field and dark images to characterize the noise of an imaging telescope between observations.
The other method uses the same data that contain the signal itself, i.e. $d_n=d_s$, which we refer to as the \emph{on-source} method, analogous to ``self-calibrating'' the data by doing noise characterization in concert with the signal processing.

\subsection{Off-source spectral estimation}

The ``off-source" spectral estimation is described in~\cite{Veitch:2014wba}.
In that case $d_n$, the data used to estimate $S_n(f)$, are data before the segment of interest that contains the signal.
In particular, $M$ nonoverlapping segments of data of duration $T$ are selected.
The data are windowed appropriately to avoid biases and the one-sided PSD is computed from each segment.
Each individual PSD exhibits large variations due to the specific noise realization in each segment.
The final noise PSD is then computed by averaging the segment PSDs by computing the median in each frequency bin.

The number of data segments $M$ needs to satisfy two criteria.
First, it needs to be large enough that the averaging can efficiently mitigate the variation due to noise realization on the resulting PSD.
Second, it needs to not be too large such that the nonstationarity of the noise becomes important.
For short-duration signals, such as BBHs, \LALInference uses a default of $M=32$, assuming there is science-quality data available for that duration prior to the segment of interest.
For longer-duration signals, such as BNSs, there is an additional constraint such that by default $M\times T \leq 1024$s.
This is chosen as a compromise between the desire for a large number of segments to average over and the avoidance of issues caused by nonstationarity.

During the initial LIGO era, the effects of nonstationary noise, the variation in the resulting off-source PSDs, and their impact on recovered binary parameters were also investigated in~\cite{Aasi:2013jjl}.
Using $M = 32$ and $T=32$s, the same simulated binary signal was added to data separated by $10$s before which an off-source PSD was estimated and then used in a PE analysis.
The spread in the recovered posterior distributions was there found to be comparable to that observed when performing the same analysis with different GW signal models between the injected ``true'' waveform and the model used to recover the signal.

\subsection{On-source spectral estimation}

The ``on-source'' spectral estimation is done with the \BayesLine algorithm originally described in~\cite{Littenberg:2014oda}, and publicly available in~\cite{bayeswave}.
\BayesLine uses a parametrized model for the noise spectrum to infer the frequency-dependent noise variance.
The model for the noise variance has two components, with the broadband noise being fit by a cubic spline where the location of control points for the spline are free parameters; while the narrow-band features in the noise are fit using a linear combination of Lorentzians, parametrized by their central frequency, amplitude, and line width.
The number of control points in the spline model and Lorentzians in the line fit are free parameters.
The model is explored using a transdimensional (or Reverse Jump) MCMC sampler~\cite{10.1093/biomet/82.4.711}. 
For a visual representation of the spectral fitting in terms of splines and Lorentzians, see Fig. 4 of~\cite{Littenberg:2014oda}.

The likelihood function used in \BayesLine has a similar form to Eq.~\ref{Lfinal} but now part of the previously neglected normalization term depends on model parameters, namely $S_n(f)$, and must be explicitly computed:
\begin{equation}\label{Lbayesline}
\mathrm{ln} {\cal{L}}(d|S_n)= -2 \sum_i^{N/2}\frac{\tilde{r}_i \tilde{r}_i^*}{ T S_n(f_i)} - \ln S_n(f_i) +{\rm const}.
\end{equation}
Note that this likelihood still implicitly assumes that the noise is stationary and Gaussian, but it relaxes the requirement on the time over which the stationary assumption must hold compared to the off-source method.

\BayesLine is fully integrated with another transdimensional MCMC algorithm, \BayesWave~\cite{bayeswave}, which uses a linear combination of wavelets to model non-Gaussian features in the data, with the option of demanding coherence across the detector network to serve as a GW signal model without relying on CBC waveforms~\cite{Cornish:2014kda}.
The combined pipeline is used for detection and characterization of short-duration ($<1$ s) GW transients in LIGO-Virgo data, including BH mergers. It possesses the flexibility to reconstruct a wider variety of signal morphologies than the dedicated CBC analyses at the expense of sensitivity to low mass and/or low signal-to-noise ratio (\SNR) events~\cite{TheLIGOScientific:2016uux,LIGOScientific:2018mvr}.

Whereas the \BayesWave pipeline directly uses \BayesLine to produces samples of the whitening filter $S_n(f)$ and marginalize simultaneously over the noise and the signal model, 
for CBC PE analyses \BayesLine has been used as a preprocessing step for spectral estimation~\cite{Abbott:2017oio,TheLIGOScientific:2017qsa,Abbott:2017gyy,Abbott:2018wiz,LIGOScientific:2018mvr}.
In that case it might be possible for the on-source spectral estimation to result in a noise model that has partly fitted the potential signal power, 
thereby corrupting the proceeding PE analyses. To prevent this from happening, the spectral estimation preprocessing step is performed with the wavelet model enabled so as to fit non-Gaussian features in the data and provide a clean residual for spectral estimation. 

While marginalizing over uncertainty in the noise model is ideal, at the time of this study, the capabilities of \LALInference require that a point estimate of the noise model is used.
When using \BayesLine for the spectral estimation application, two options for point estimates are available: The ``Fair Draw'' model, which is the reconstructed $S_n(f)$ taken from a single random sample of the Markov chain; or the ``Median'' model, which is assembled by taking the median value in each frequency bin of the posterior distribution of $S_n(f)$.
In this paper we will test the performance of both approaches, but ultimately lean in favor of the median model thanks to its reproducibility.

\BayesLine uses a fairly informative model about the noise spectrum i.e. a smoothly varying function, described by cubic splines, with prominent
spikes described with Lorentzians. This model has been informed by past experience on the behavior of typical LIGO noise spectra. At the same 
time the priors on the various model parameters are less informative:
uniform priors on the spline parameters (location in [$f_{\rm spline},\log S_n]$ space of the spline points) and line parameters (frequency, natural log of the amplitude, and line width).
These uniform priors do not take full advantage of our knowledge about the data--the model has to ``re-learn'' where spline points and lines are needed every time it analyzes new data despite there being persistent features in the noise spectrum throughout an observing run, e.g. general line locations.
It is a desirable improvement of the noise modeling approach to develop priors from the data that are continually adapted throughout an observing run, giving the on-source spectral estimation method the ``best of both worlds'' where it is informed by long strides of data such as the off-source method, but preserves the minimal requirements on the stationary timescales of the noise.
We leave these developments to future work, though we emphasize that the current implementation of \BayesLine is still able to produce
converging posteriors for the noise model $S_n(f)$. With the typical number of spline points and lines, the posterior on the PSD is likelihood dominated, so more informative priors are only expected to reduce the computational cost and overall
timescale of the analysis. More details will be described elsewhere~\cite{BLMethodsII}.

\subsection{Injections}
\label{inj}

In order to study the above methods of estimating the noise variance of real interferometric detector data, we use data from the second observing run (O2) of the advanced detectors accessible through the Gravitational Wave Open Science Center~\cite{GWOSC,Vallisneri:2014vxa}.
We simulate CBC signals, add (or ``inject'') them to the observational data, and then analyze the data using the same PE procedures used in Ref.~\cite{LIGOScientific:2018mvr}.
Results from the simulated CBC signals are grouped into three types: high mass BBHs, low mass BBHs, and BNSs.
The PE analyses use the publicly available software library \LALInference to sample the multidimensional posterior distribution of the source parameters~\cite{Veitch:2014wba}.
Details of the injections are as follows:

\begin{enumerate}
\item High mass BBH: We draw $38$ random samples from the posterior distribution for GW150914~\cite{O1catalogPErelease}, create simulated GW signals using the spin-precessing waveform model IMRPhenomPv2~\cite{Husa:2015iqa,Khan:2015jqa,Hannam:2013oca}, and inject those signals into non-ovelapping data segments around the GW event GW170104.
We then analyze $4$s of data containing each simulated signal, over the bandwidth from 20 to 1024 Hz.
We use again IMRPhenomPv2, in its reduced order quadrature implementation~\cite{Smith:2016qas}, to recover the properties of the signal.
\item Low mass BBH: We follow the same procedure as the high mass BBH injections
above, only we now use $22$ random samples from the posterior distribution for
GW151226~\cite{O1catalogPErelease}. The analysis segment is also increased to
$8$s.  We keep the bandwidth the same.
\item BNS: We use $20$ randomly selected samples from the posteriors computed in Ref.~\cite{Chatziioannou:2018vzf} for simulated BNS signals with different models for the NS equation of state.
In particular, we use 7(7)[6] samples from the WFF1(H4)[MS1] posteriors, allowing us to probe a wide range of equation of state stiffness.
We create simulated GW signals using the spin-aligned waveform model IMRPhenomD\_NRTidal~\cite{Dietrich:2018uni}, and inject the signals in nonoverlapping segments of O2 data.
We do not vet the data we use for glitches before the analysis.
We then analyze $128$s of data containing each simulated signal, in a bandwidth of $25$ to $2048$Hz.
We use again IMRPhenomD\_NRTidal, in its reduced order quadrature implementation~\cite{Smith:2016qas,smith_rory_2019_3255081}, to recover the properties of the signal.

\end{enumerate}

In all cases, we estimate the on-source noise variance after the signals have been injected in the data, as is true for real detections. Additionally,
all PE analyses use the same prior distributions as~\cite{LIGOScientific:2018mvr}.

\section{Whitening tests}
\label{tests}

\begin{figure*}[]
\includegraphics[width=.9\columnwidth,clip=true]{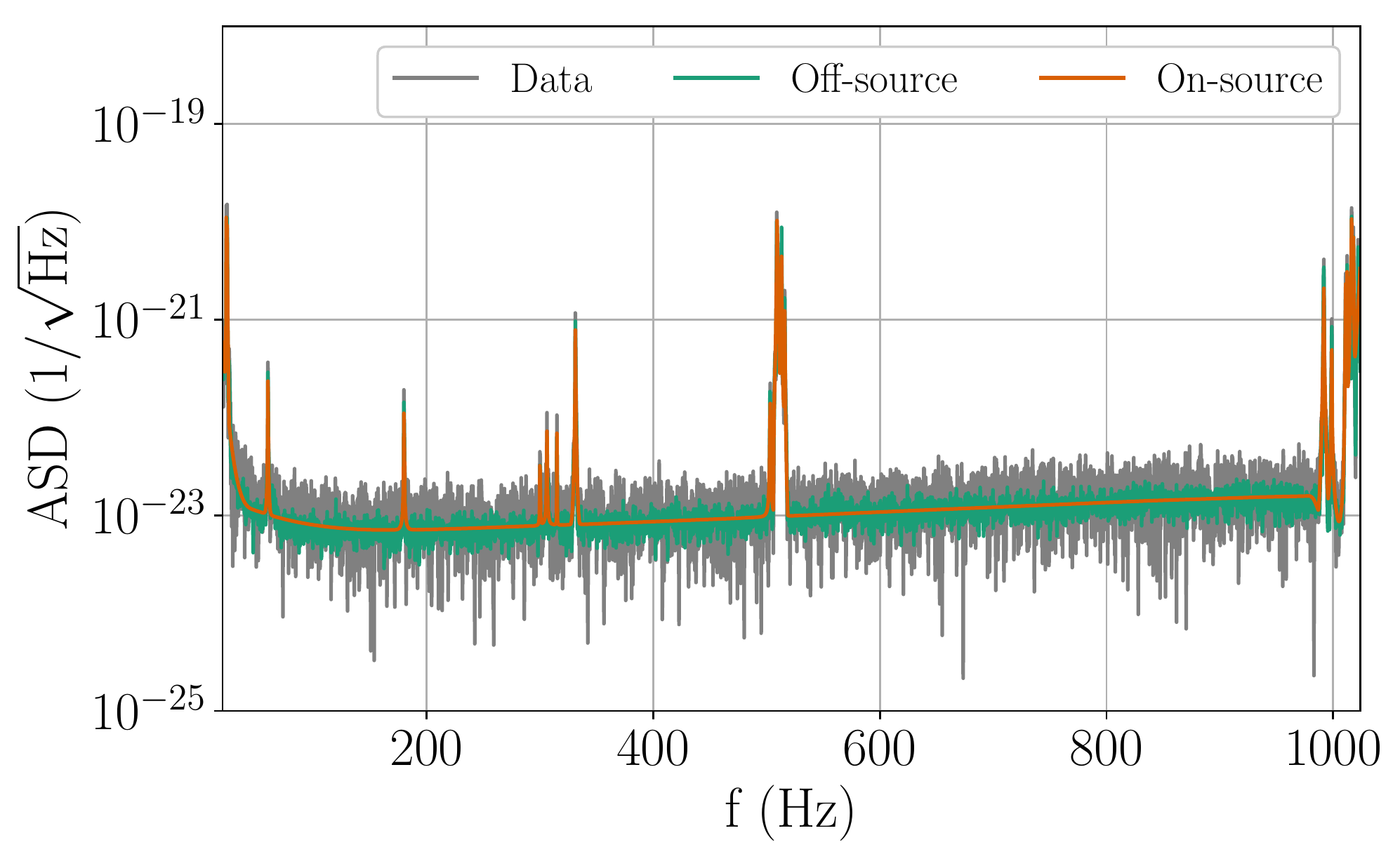}
\includegraphics[width=.9\columnwidth,clip=true]{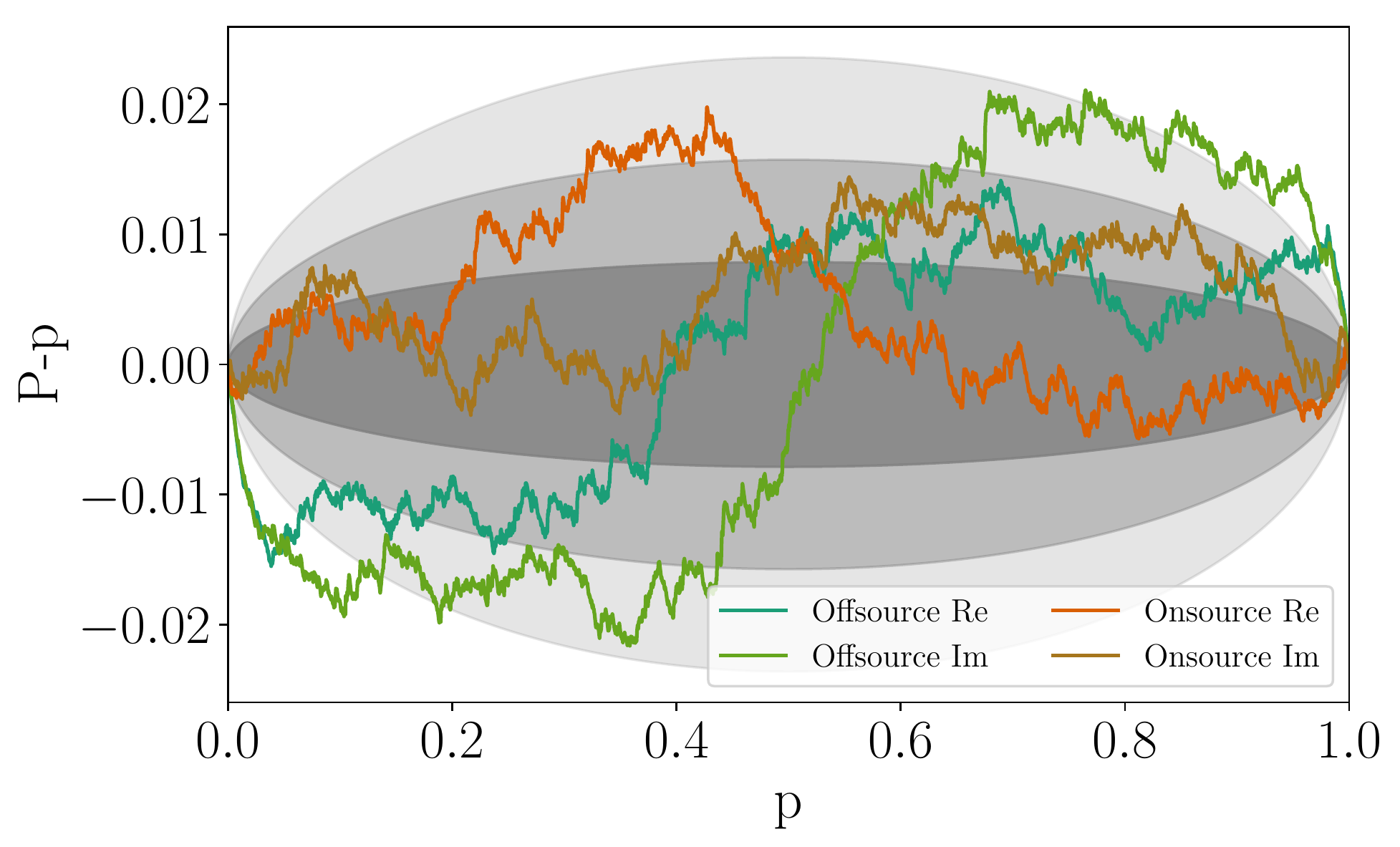}\\
\includegraphics[width=.9\columnwidth,clip=true]{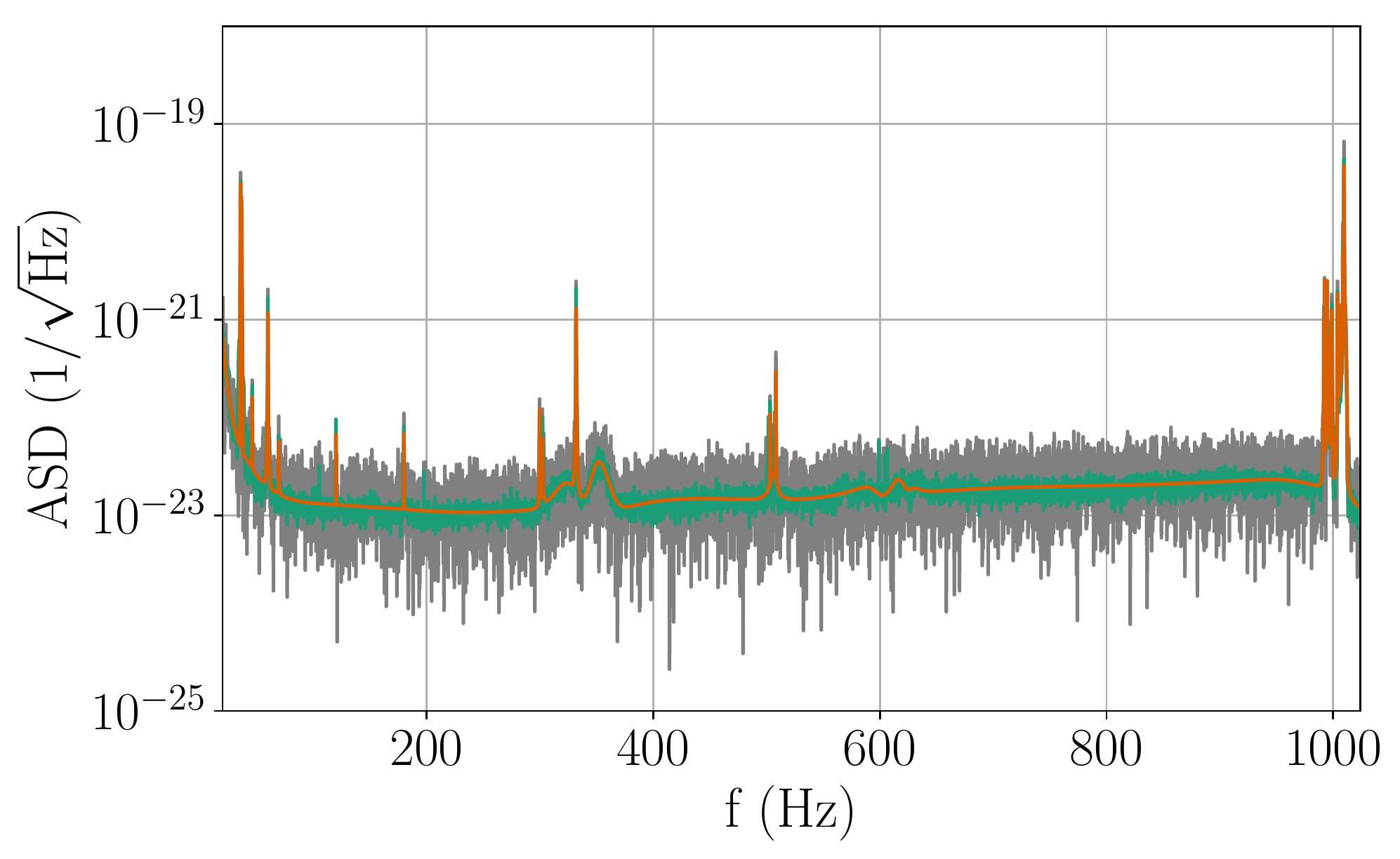}
\includegraphics[width=.9\columnwidth,clip=true]{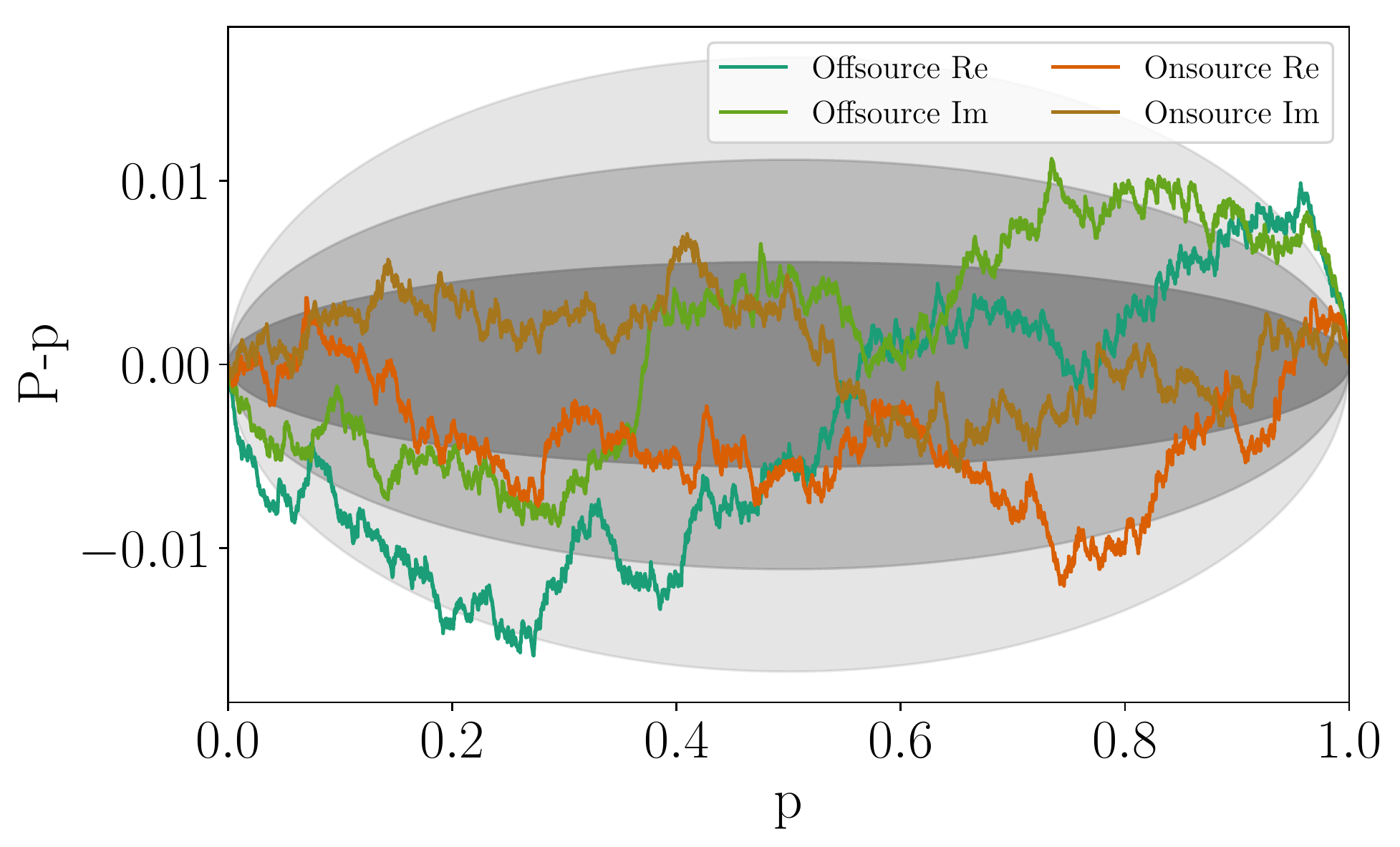}\\
\includegraphics[width=.9\columnwidth,clip=true]{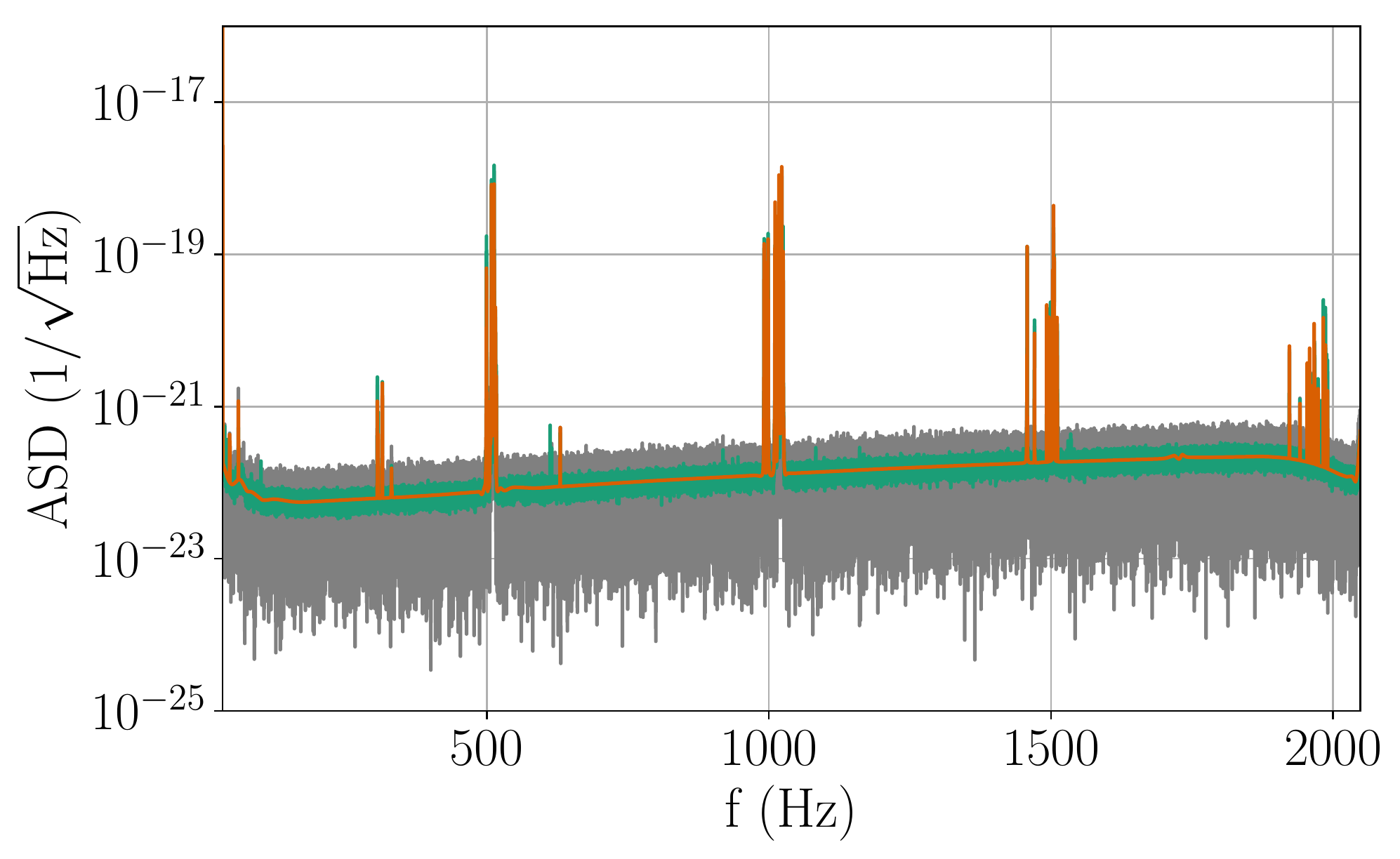}
\includegraphics[width=.9\columnwidth,clip=true]{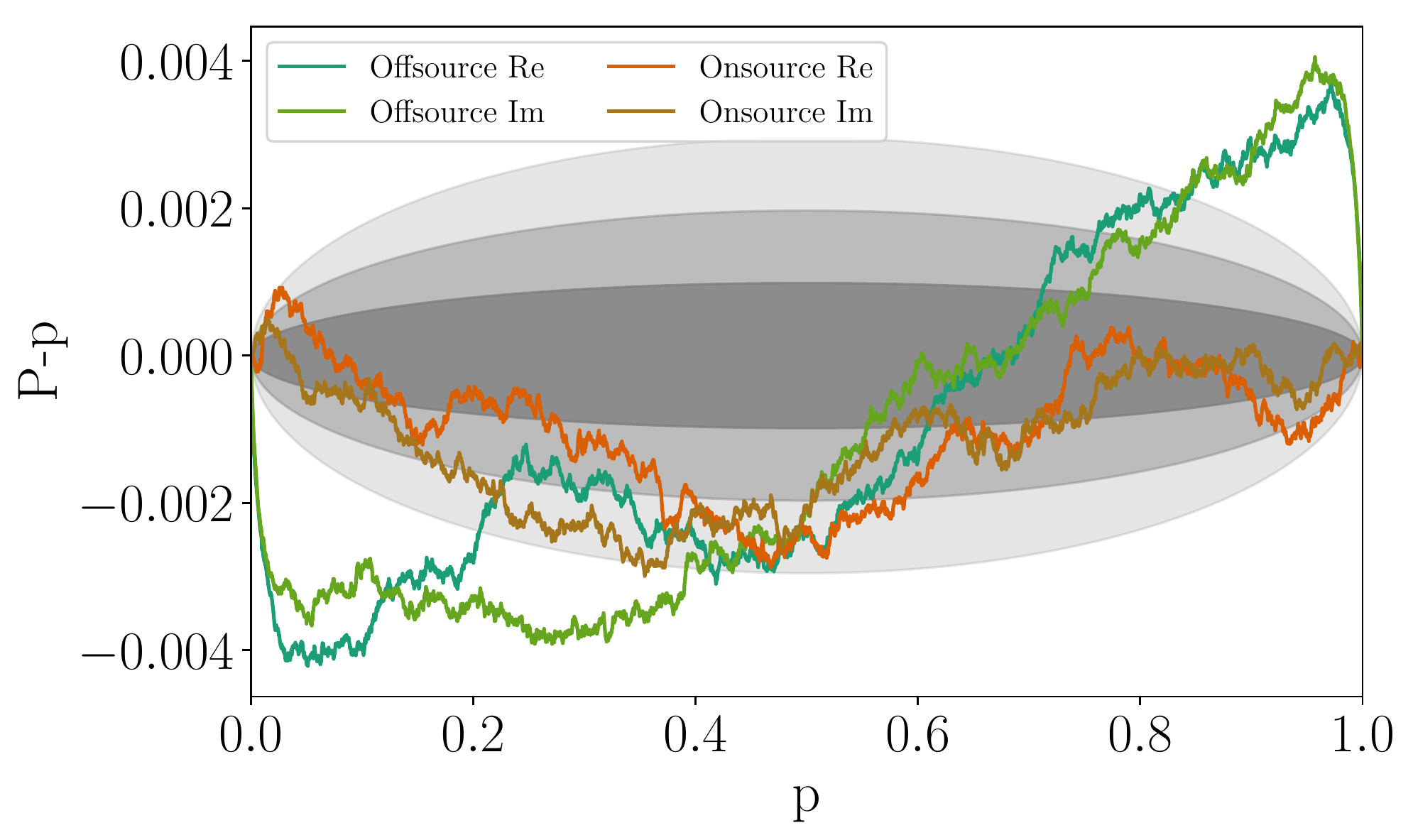}
\caption{Example results for a high mass BBH system (top), low mass BBH system (middle), and BNS system (bottom).
The left column shows the amplitude spectral density of the interferometer data plus signal injection (gray), the off-source spectral estimation (green), and the median on-source noise model (orange).
The right column shows the difference between the cumulative distribution of the percentiles ($P$) and expected credible level ($p$) versus the expected credible level ($p$ ) of the whitened Fourier amplitudes, assuming they are drawn from $\normal$.
The gray ellipses represent expected fluctuations at the $1$, $2$, and $3\sigma$ level given the number of samples in the data.}
\label{fig:PSDs}
\end{figure*}

Having calculated the noise variance of the data around the injections described in Sec.~\ref{inj} using the on- and off-source methods, we perform a number of tests to study the performance of the spectral estimation in the context of the assumptions implicitly made by the likelihood function used within GW PE.
Figure~\ref{fig:PSDs} shows representative results from analysis of data containing high mass BBH (top row), low mass BBH (middle row), and BNS (bottom row) injections.

The left column shows a comparison of the amplitude spectral density (ASD) of the on-source data that are being analyzed (grey), the off-source whitening filter (green), and the median on-source whitening filter (orange).
The ASD of the data has a large variance since it is affected by the specific noise realization.
The off-source spectral estimate has lower variance obtained by averaging over many segments of equal length.
The on-source parametrized spectral model has the least degrees of freedom, and imposes a smooth fit through its construction, resulting in the least variation across adjacent frequency bins.

A visual comparison between the spectral estimates reveals that the off-source model contains a small number of 
low-amplitude spectral lines that are not obviously present in the power spectrum of the data, and do not appear in the on-source model for the spectrum.
One such example can be found at around 600Hz in the middle row.
These additional lines rise above the broadband noise by a factor of a few.
They can be attributed to the fact that the ``strength'' of spectral lines grows with duration of the data $T$ (as the $\sqrt{T}$ in amplitude, $T$ in power).
For short data segments there is insufficient evidence for low-amplitude lines to be included in the \BayesLine fit.
The off-source PSD estimation approach uses more data, and therefore \emph{is able to accumulate} evidence for lines not necessarily prominent in the shorter segment of data being used for PE, but at the expense of increased demand on the stationary timescales of the noise.
We emphasize that Fig.~\ref{fig:PSDs} directly plots the data that enter the numerator of the likelihood function (in grey) and hence provides the most direct comparison in favor of the presence or absence of weak lines.
For completeness, we have checked that all lines appearing in the on-source model are also visually present in the power spectrum of the data.

Finally, as a word of caution we note that the prospect of ``missing lines'' that are known to be in the data from longer integration times sounds alarming, but the GW signal to noise ratio (and, more importantly, the likelihood) for CBC sources is integrated over a large range of frequencies and these weak lines occupy a small fraction of the overall observing bandwidth.
The integrated differences from the broadband noise when using off-source estimates are also important to consider when comparing PSD estimates.

The right column of of Fig.~\ref{fig:PSDs} shows the distribution of the whitened residuals for each of the spectral estimates on the left plots.
We compute the plot in the following way.
We begin with the complex frequency-domain data, the ASD of which is plotted in the left column of the figure.
We then use the estimated noise variances (orange and green curves in the left column) as whitening filters, dividing the real and the imaginary part of the data by the square root of the noise model.
If the assumptions about the noise statistics are valid, and the whitening filter is a good approximation to the true noise variance, then the real and the imaginary part of the whitened data should be consistent with random draws from $\normal$, a zero-mean, unit-variance normal distribution.
We test noise assumptions for each whitened data point by computing its corresponding percentile of $\normal$ which, in turn, should be uniformly distributed between $0$ and $1$ if the null hypothesis (the noise is stationary and Gaussian etc.) is supported by the data.
We then compute the cumulative distribution function of the percentiles, i.e. the number of percentiles $P$ that are are below a certain value $p$.
Because of the large number of data points, we plot $P-p$ as a function of $p$ (rather than $P$ vs. $p$) to assess if the data and noise model satisfy our assumptions in the likelihood function.
The grey shaded regions enclose the $1$, $2$, and $3\sigma$ expected variation of $P-p$ given the finite number of samples in the data.
The latter is computed through $\sigma^2=p(1-p)/N$, where $N$ is the number of data points.

For independent samples drawn from a normal distribution, the above procedure should produce $P-p$ values that are close to $0$.
We find that indeed the on-source spectral estimation method leads to whitened data whose $P-p$ values are within the expected variation.
The off-source method, on the other hand, results in large outliers, i.e. data points that would be unlikely random draws from a Gaussian distribution, and the $P-p$ distribution ventures outside of the expected range more significantly than the on-source method.
The BNS example shows the largest discrepancy between the spectral estimation methods, and from the null hypothesis.
Analysis of BNS signals requires the longest data segments, which puts the assumptions about stationarity under pressure both for the on-source method (now using $128$s of data rather than $4$ or $8$s for the BBH cases), and even more so for the off source method which needs to average over several $128$s-long segments of data.
It is therefore expected that the differences in spectral estimation approaches will be most apparent for BNS analyses.

\subsection{The Anderson-Darling statistic}

\begin{figure}[]
\includegraphics[width=.9\columnwidth,clip=true]{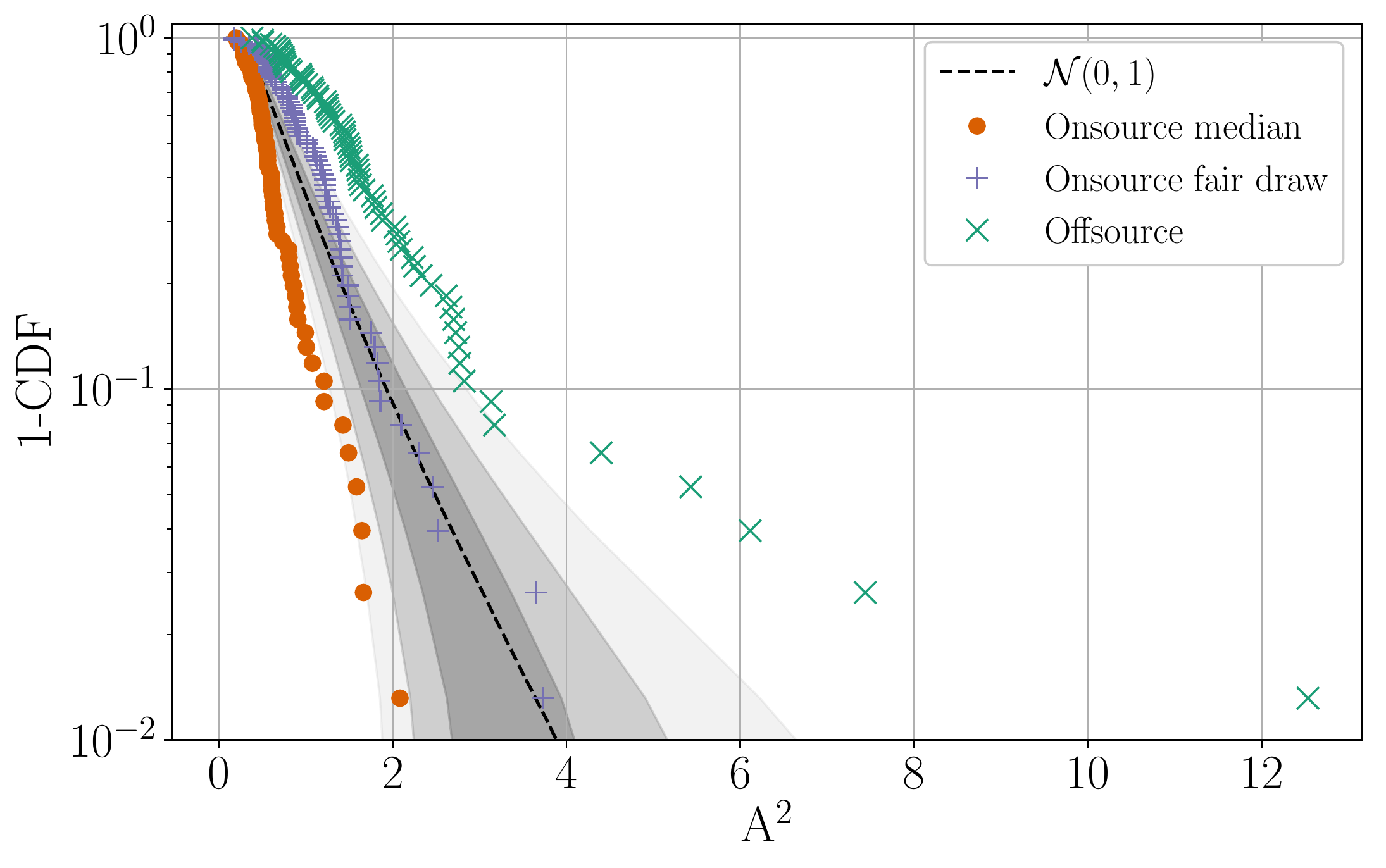}\\
\includegraphics[width=.9\columnwidth,clip=true]{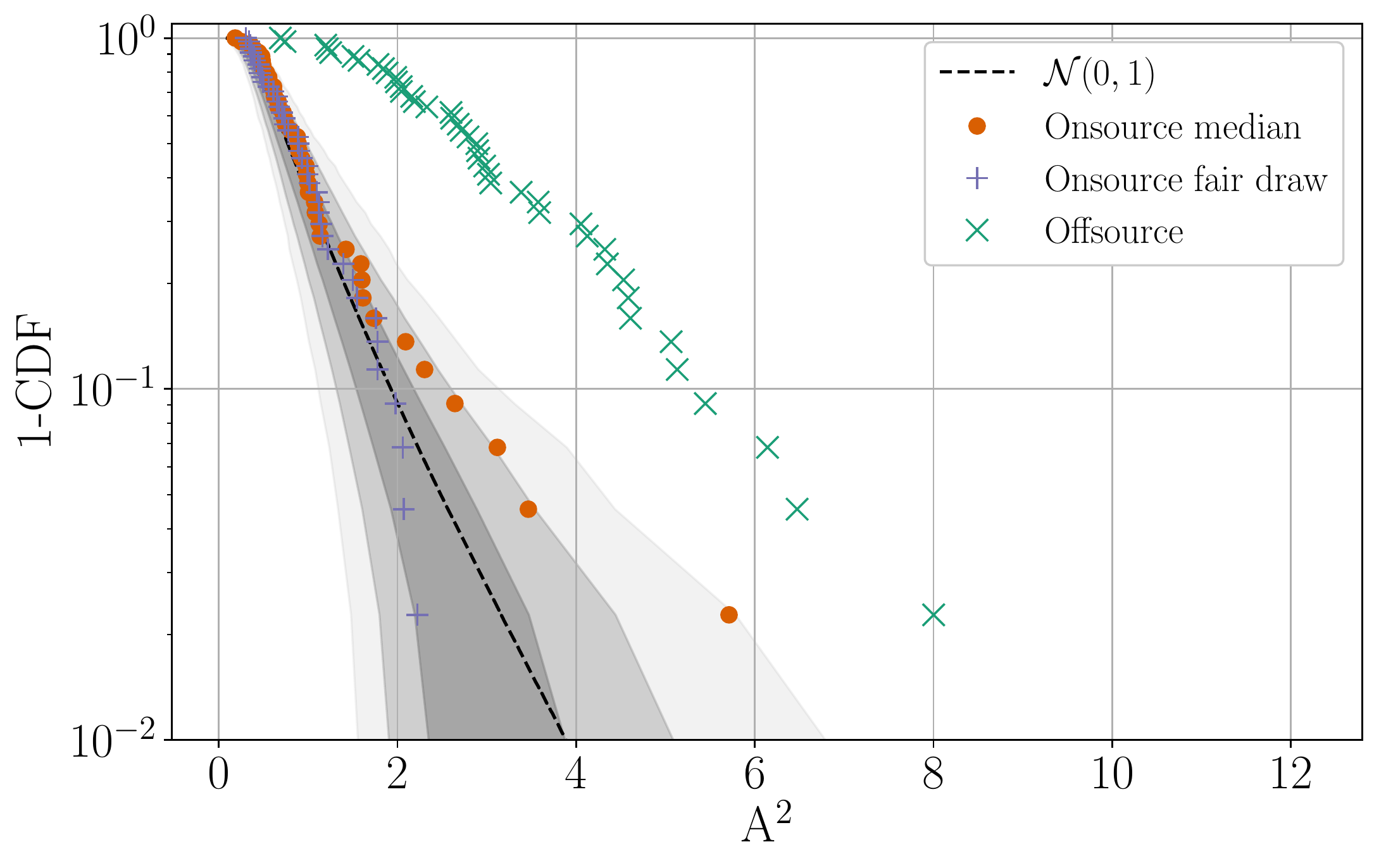}\\
\includegraphics[width=.9\columnwidth,clip=true]{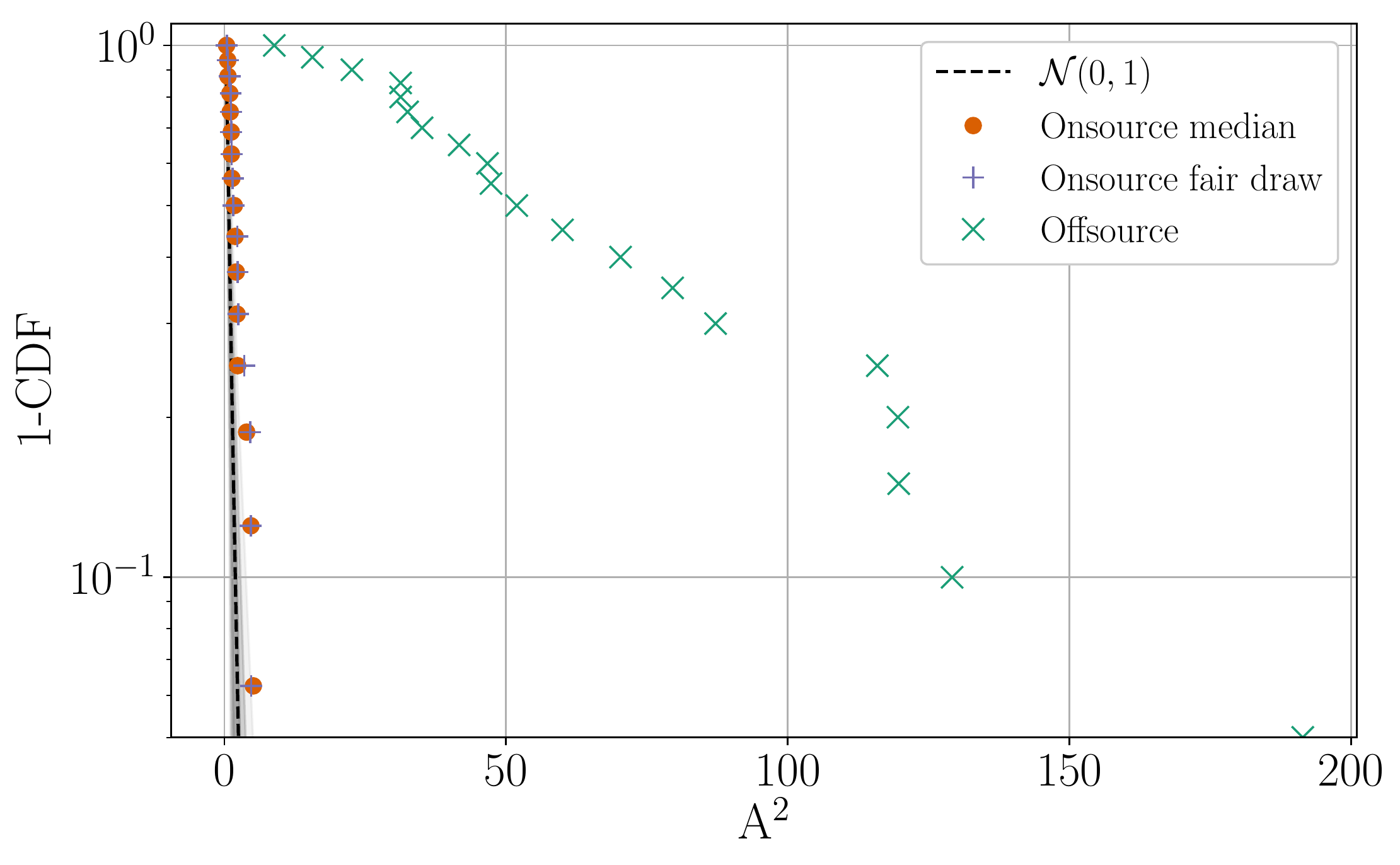}
\caption{Cumulative distribution of the square of the AD statistic for our high mass BBH (top), low mass BBH (middle), and BNS (bottom) injections.
The black dashed line shows the theoretical expectation for random samples drawn from a normal distribution with a mean of zero and a
standard deviation of one. The grey shaded regions show $1, 2$ and $3\sigma$ error regions.}
\label{fig:ADAll}
\end{figure}

The previous section showed representative examples of the different spectral estimation methods, and a visual demonstration of how the quality of the data whitening depends on the noise model used for the analysis.
In this section we use the entire injections set to quantify the differences in spectral estimation methods for an ensemble of sources and data segments.
To do so, we need a statistic for characterizing the quality of the data whitening for each analysis segment.
In this study we adopt the Anderson-Darling statistic $A^2$, which is a way of assessing whether a set of samples are drawn from a given probability distribution, in this case $\normal$.
It is also related to the $p$-value of the hypothesis that the samples are drawn from the target distribution. 
In general, samples that are inconsistent with the target distribution lead to large $A^2$ values and low $p$-values. This in turn means that 
we can reject the hypothesis that the samples were drawn from the target distribution.

It is defined as~\cite{AndersonDarling}
\begin{equation}
A^2 \equiv N \int_{-\infty}^{\infty} \frac{ (F_n(x)-F(x))^2 }{F(x)(1-F(x))}d F(x)
\end{equation}
where $N$ is the number of samples, $F(x)$ is the target distribution -- in this case $\normal$ -- and $F_n(x)$ is the empirical distribution of the samples.
The test is a measure of the integrated distance between $F(x)$ and $F_n(x)$ computed with the metric $F(x)(1-F(x))$.
The latter is nonunique; this specific choice of metric places more weight on the tails of the observed distribution.
This is appropriate for GW PE, as GW signals are observed as outliers of the expected noise distribution.

\begin{figure}[]
\includegraphics[width=.9\columnwidth,clip=true]{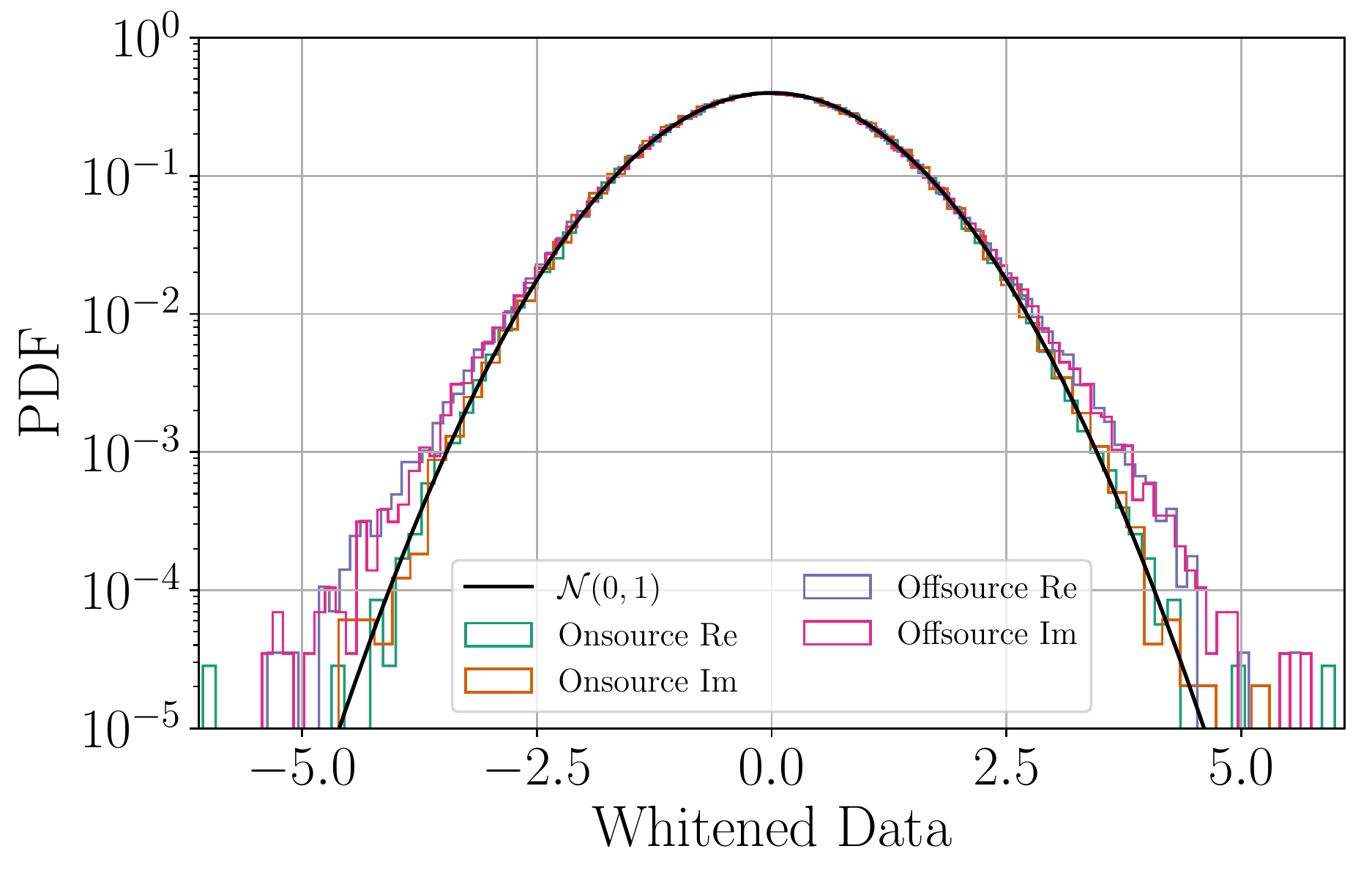}
\caption{Histogram of the real and the imaginary part of the whitened data obtained with the on-source and the off-source whitening filter for the same system as the bottom row of Fig.~\ref{fig:ADAll}. The x-xis is in units of standard deviation.
For comparison we also plot a zero-mean, unit-variance Gaussian distribution.
The off-source whitening is done on the data before the simulated injections are added.
The on-source whitening has the signals included, but \BayesWave\ fits any non-Gaussian noise with wavelets and we whiten the residuals.
}
\label{fig:BNShist}
\end{figure}

For each of our injected signals, we compute $A^2$ using different spectral estimates as whitening filters, and plot the cumulative distributions in Fig.~\ref{fig:ADAll}.
In this figure we include both the on-source median and fair draw models, as well as the results using the off-source method.
The black dashed lines show the expected distribution of $A^2$ empirically determined by 10000 Monte Carlo realizations of $\normal$, while the grey shaded regions show the $1, 2$ and $3\sigma$ error regions.
We use the results of~\cite{JSSv009i02} to compute the corresponding $p$-values and confirm that the $p$-values for a given value of $A^2$ that would be inferred from this distribution are consistent with the theoretical expectation.

We find that the off-source filter frequently produces whitened data with low probability of being generated by $\normal$, i.e. whitened data that strain the assumptions implicit in the likelihood function used by PE.
This suggests that the data whitened with the off-source filter have multiple outliers from $\normal$ that cannot simply be explained as a random variation.
As an example of this, in Fig.~\ref{fig:BNShist} we plot the histograms of the real and the imaginary Fourier amplitudes of the whitened data with the on-source and the off-source filter and compare them to $\normal$.
The example corresponds to the same system as the bottom row of Fig.~\ref{fig:ADAll}.
This system has $A^2=51$ with the off-source filter, and as expected we see that the data histogram deviates from a normal distribution at around $3\sigma$ or earlier.

From Fig.~\ref{fig:ADAll} we also see that the on-source models lead to $A^2$ values that are within the expected variation (grey regions) and hence more consistent with the null hypothesis than the off-source method.
For high mass BBHs (top row, using 4 s of data) we find that the best agreement between the fair draw and the theoretical expectation, while the median model yields high $A^2$ results at a lower rate than expected, though still consistent with the expected distribution at $3\sigma$.
Lower values of $A^2$ than expected suggest that the whitened data exhibit \emph{fewer} outliers from Gaussianity than random numbers produced by a random number generator.
We attribute this to the fact that the analysis segment is short and the spectral lines are not well resolved.
Both the median and the fair draw on-source filters follow the theoretical expectation for low mass BBHs with good accuracy, while for BNSs we find that the on-source filters outperform the off-source one by a wide margin.
This behavior is also reflected in Fig.~\ref{fig:BNShist} where we see that the data whitened with the on-source filter follow the normal distribution curve to at least $4\sigma$, where low-sample variation takes over; the corresponding AD value is $A^2=4.5$.

We emphasize that differences between off- and on-source methods are increasingly pronounced when more data are used in the analysis, therefore placing a larger demand on the noise being stationary over the interval needed for off-source spectral estimation.
Though the obtained $A^2$ distribution for BNS (bottom row) is consistent with the theoretical expectation to within a $3\sigma$ uncertainty, we do not expect this to be the case for longer analysis segments if the trend continues.
Such longer data segments will soon become unavoidable as the lower frequency performance of the detector improves towards design sensitivity or next generation detectors~\cite{2010CQGra..27h4007P,2011CQGra..28i4013H}.
Similar challenges will be faced by analysis of data from the planned space-based detector LISA~\cite{LISAproposal,Edlund:2005}.

\section{Parameter Estimation}
\label{PE}

The statistical tests of the whitened data presented in the previous section confirm that the on-source spectral estimation results in whitened data
that more closely follow the requirements of a Gaussian likelihood. In this section, we study how the deficiencies of the off-source spectral estimation method  
affect the inferences made about the GW signals.
To assess how the noise model affects PE we analyze each injected signal with the publicly available software library \LALInference~\cite{Veitch:2014wba} to obtain samples from the posterior distribution of the system parameters using the off-source, on-source median, and on-source fair draw noise models.
To reduce computational cost, and since we have established similar performance between the on-source median and fair draw models, we only compare the off-source model to the on-source median model in our analyses of the BNS signals.

\begin{figure}[]
\includegraphics[width=.9\columnwidth,clip=true]{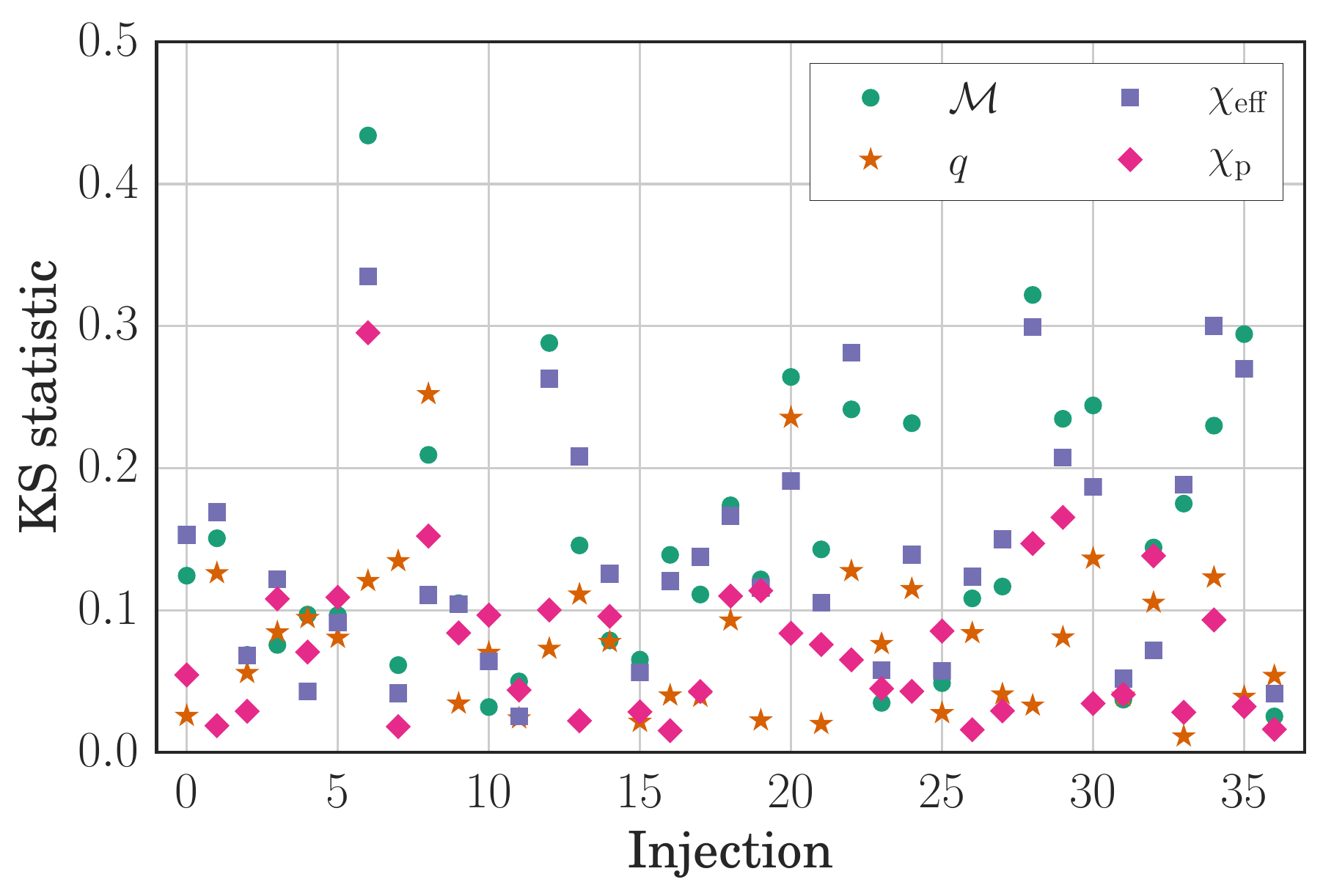}\\
\includegraphics[width=.9\columnwidth,clip=true]{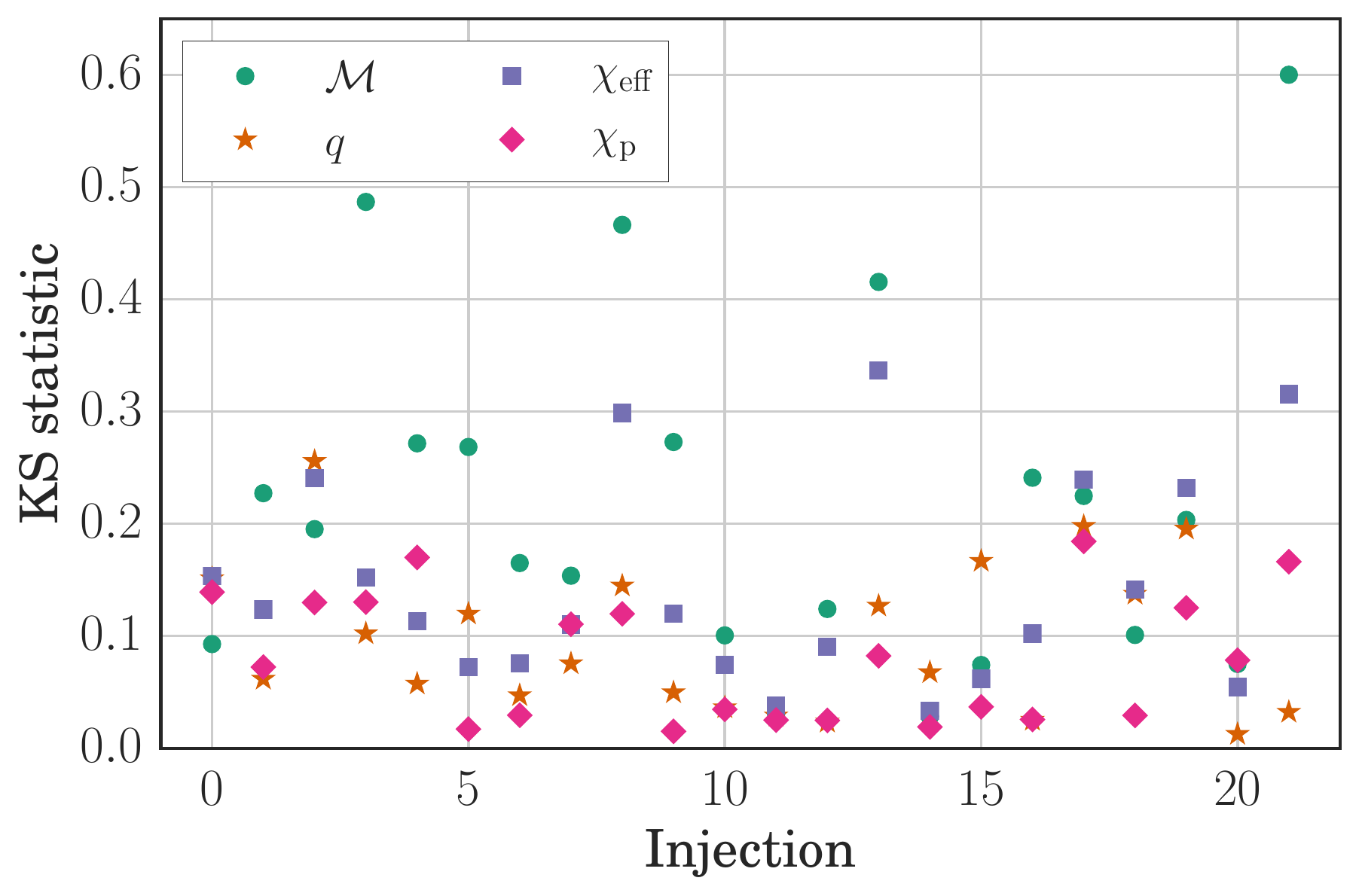}\\
\includegraphics[width=.9\columnwidth,clip=true]{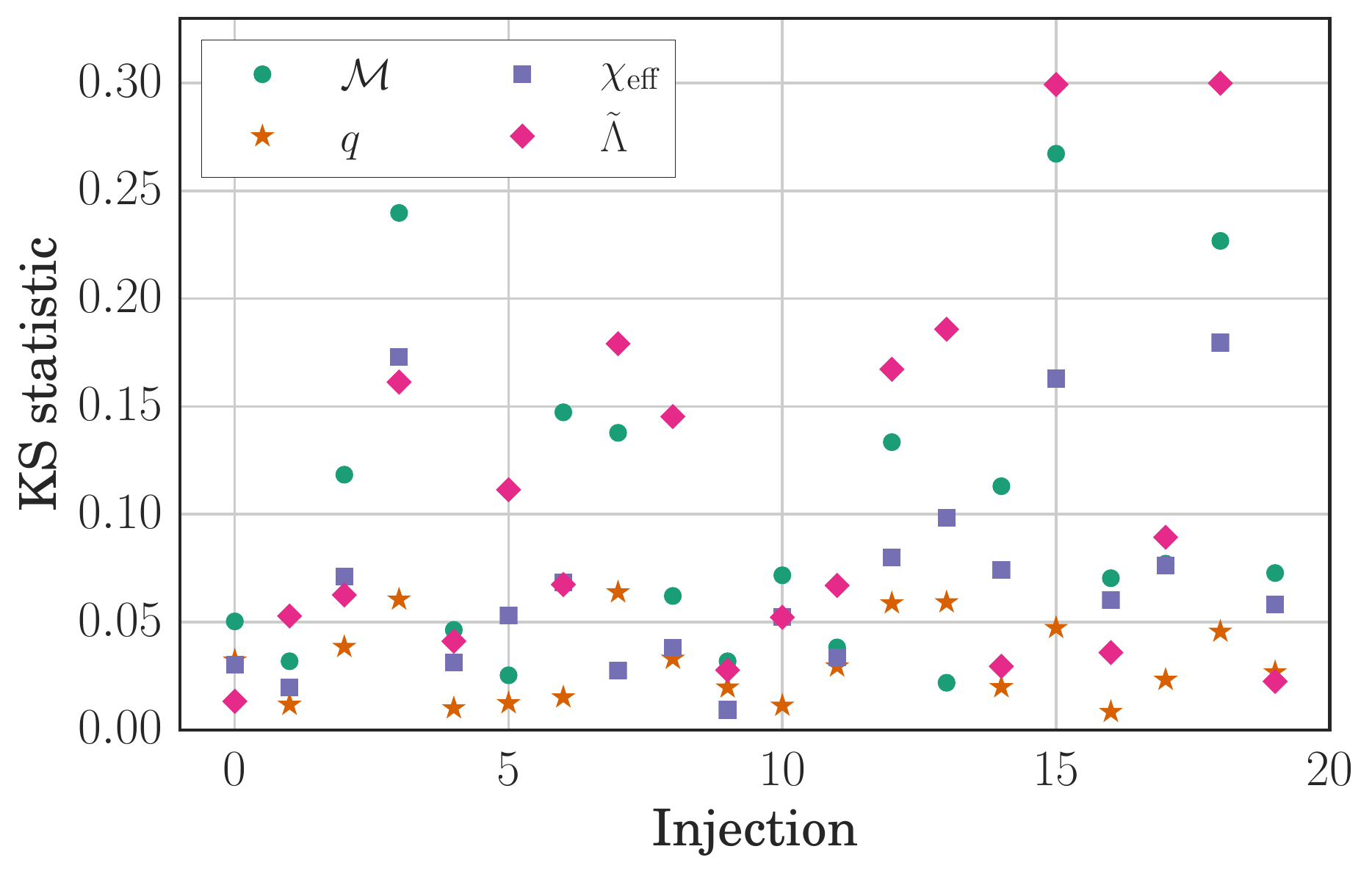}
\caption{KS statistic between the posteriors for selected parameters computed with the off-source and the on-source median noise variance for our high mass BBH (top), low mass BBH (middle), and BNS (bottom) injections.}
\label{fig:KSAll}
\end{figure}

We quantify the difference between the posteriors for the various system parameters using the Kolmogorov-Smirnov (KS) statistic.
The KS statistic is defined as the maximum distance between two cumulative distributions, in this case the one-dimensional posterior for a given parameter obtained with the on-source or the off-source noise models.
Figure~\ref{fig:KSAll} shows the KS statistic for selected parameters of each of our injections when using the off-source median method. We find that the on-source median and fair-draw methods produce very similar results compared to the off-source method.

We find that in general the largest KS statistic, and hence the largest difference between the posterior estimates, is obtained for the chirp mass, defined as $\mathcal{M} = (m_1 m_2)^{3/5}/(m_1+m_2)^{1/5}$.
This is likely due to the fact that the chirp mass is the best measured mass parameter, and hence is sensitive to systematic errors due to, e.g., unsatisfied assumptions in the data model (i.e. likelihood function).
The ratio of the masses of the two binary components $q=m_2/m_1, \; (m_1 > m_2)$ is less well measured, so it is less affected by the method for estimating the noise.
The effective spin parameter, defined as the mass-weighted projection of the spin components along the orbital angular momentum~\cite{Racine:2008qv}, also results in moderately large KS statistic for some injections.
For the spin-precessing BBH injections, we also present results for the effective precession parameter $\chi_p$~\cite{Schmidt:2014iyl}.
Finally, for the BNS injections, we examine the effective tidal parameter $\tilde{\Lambda}$~\cite{Wade:2014vqa}, which is the best measured tidal parameter.
The remaining figures of this section show example posterior distribution for the injections with the largest KS statistic.

\begin{figure}[]
\includegraphics[width=\columnwidth,clip=true]{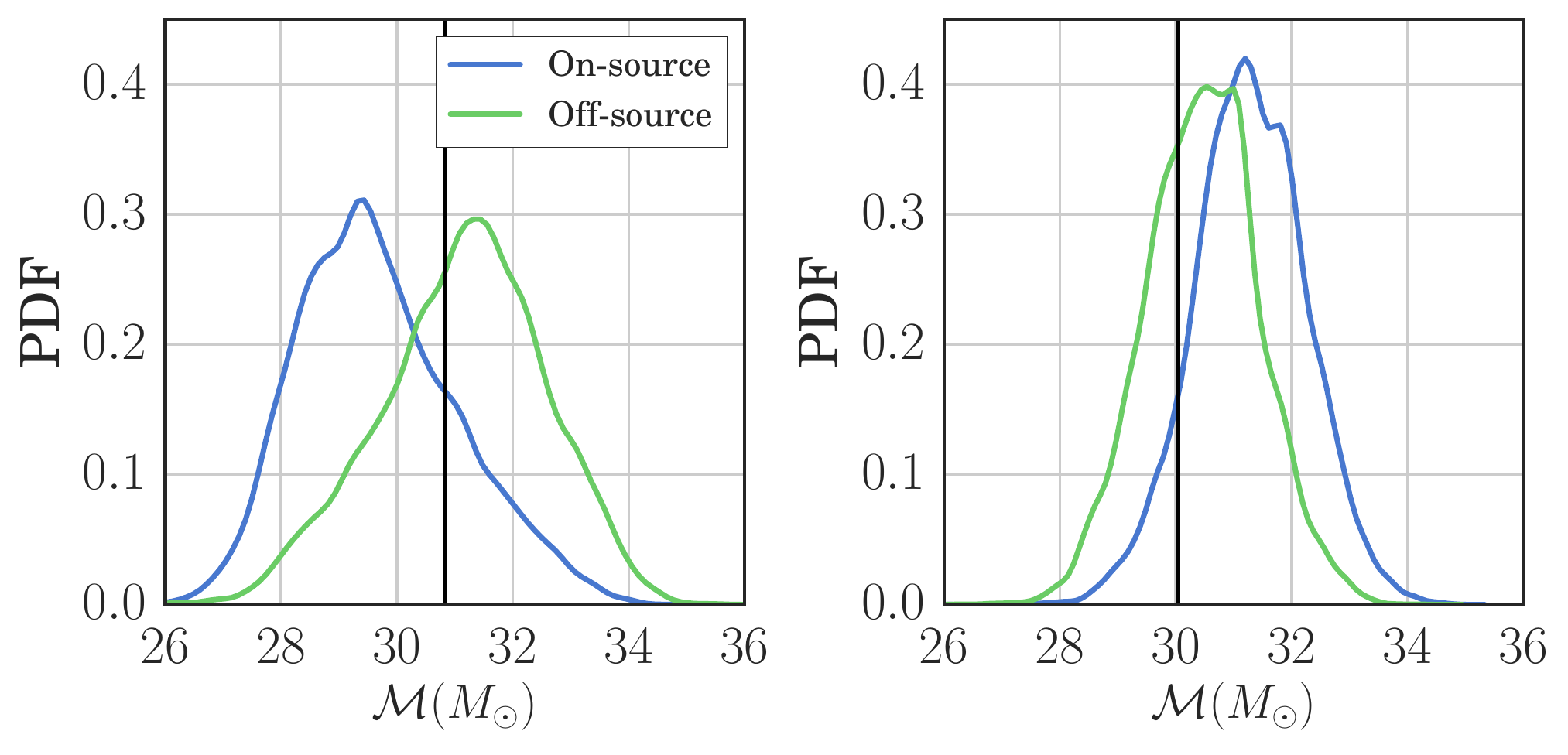}\\
\includegraphics[width=\columnwidth,clip=true]{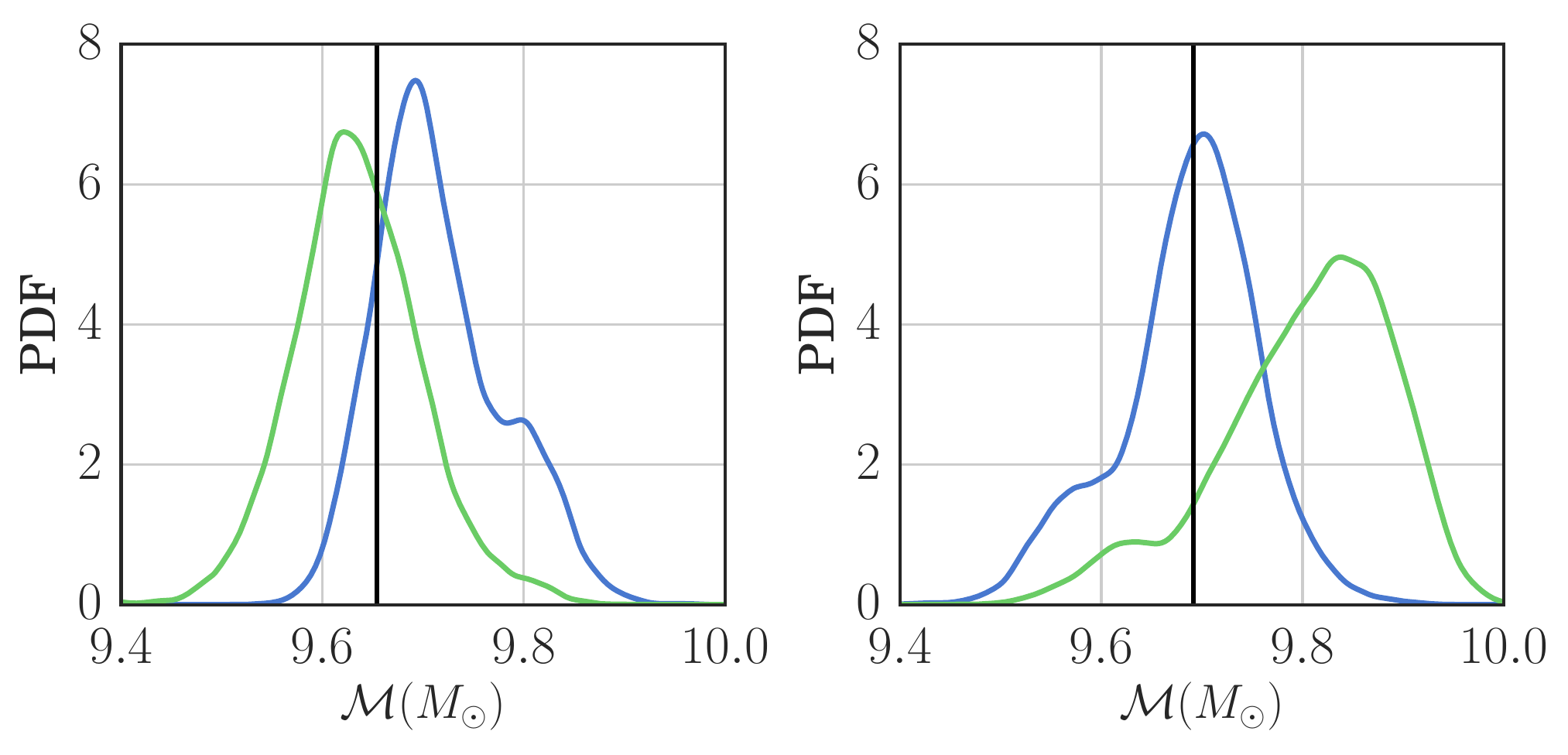}\\
\includegraphics[width=\columnwidth,clip=true]{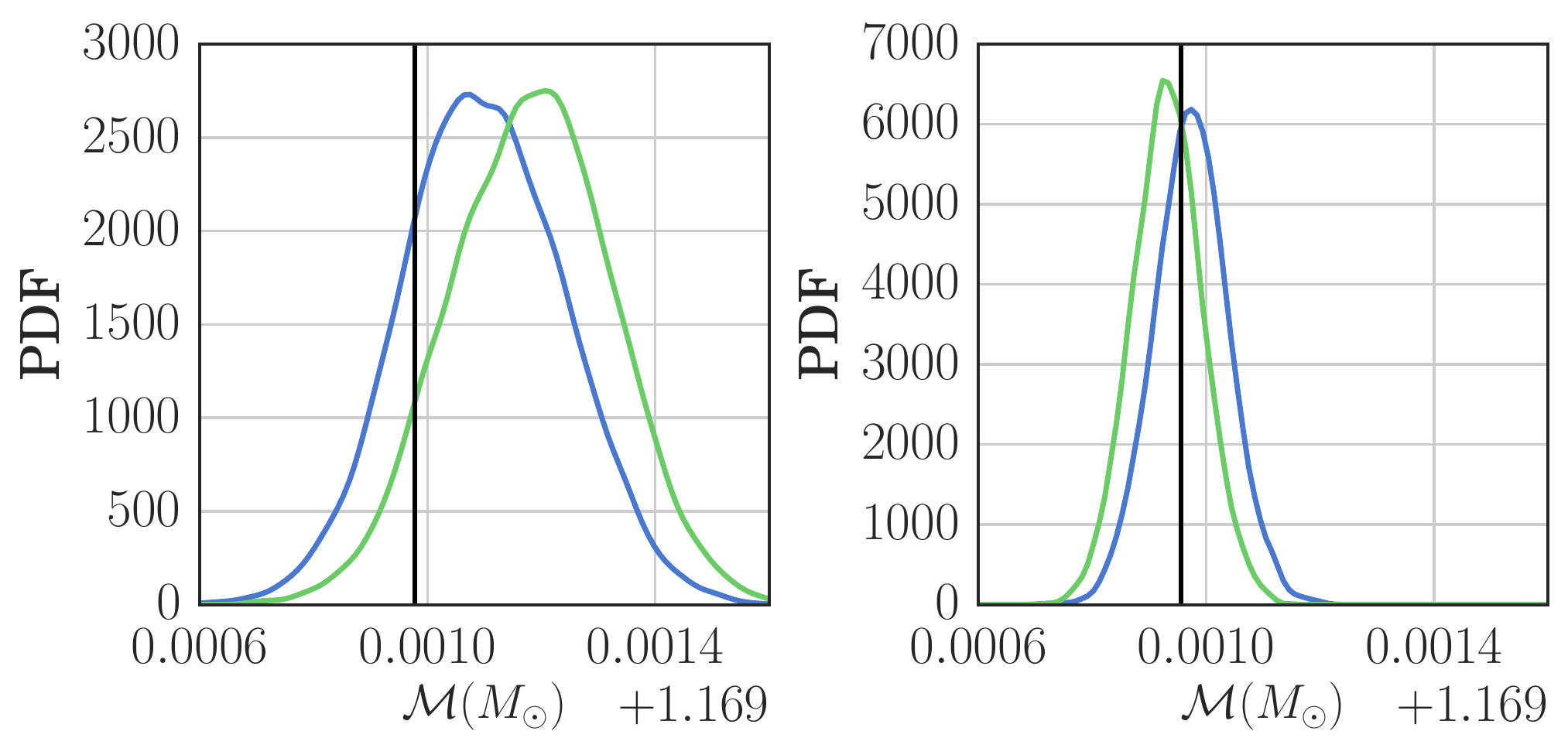}
\caption{Posterior distributions for the chirp mass obtained with the off-source and the on-source whitening filter for the two high mass BBH (top), low mass BBH (middle), and BNS (bottom) systems with the highest KS statistic from Fig~\ref{fig:KSAll}.
The true chirp mass is represented by the black vertical line.}
\label{fig:McAll}
\end{figure}

We begin in Fig.~\ref{fig:McAll} with the chirp mass distributions obtained with the off-source and on-source median noise variance for our different sets of injections.
For reference, the vertical black lines are the injected values of the chirp mass.
We do emphasize though that the one-dimensional posterior distribution for a parameter is not generically expected to peak at the injected value, especially in the presence of noise.
As expected, we find that the chirp mass is relatively well measured, so a small difference in the estimate of the noise can have a visible impact on the resulting posteriors.

\begin{figure}[]
\includegraphics[width=\columnwidth,clip=true]{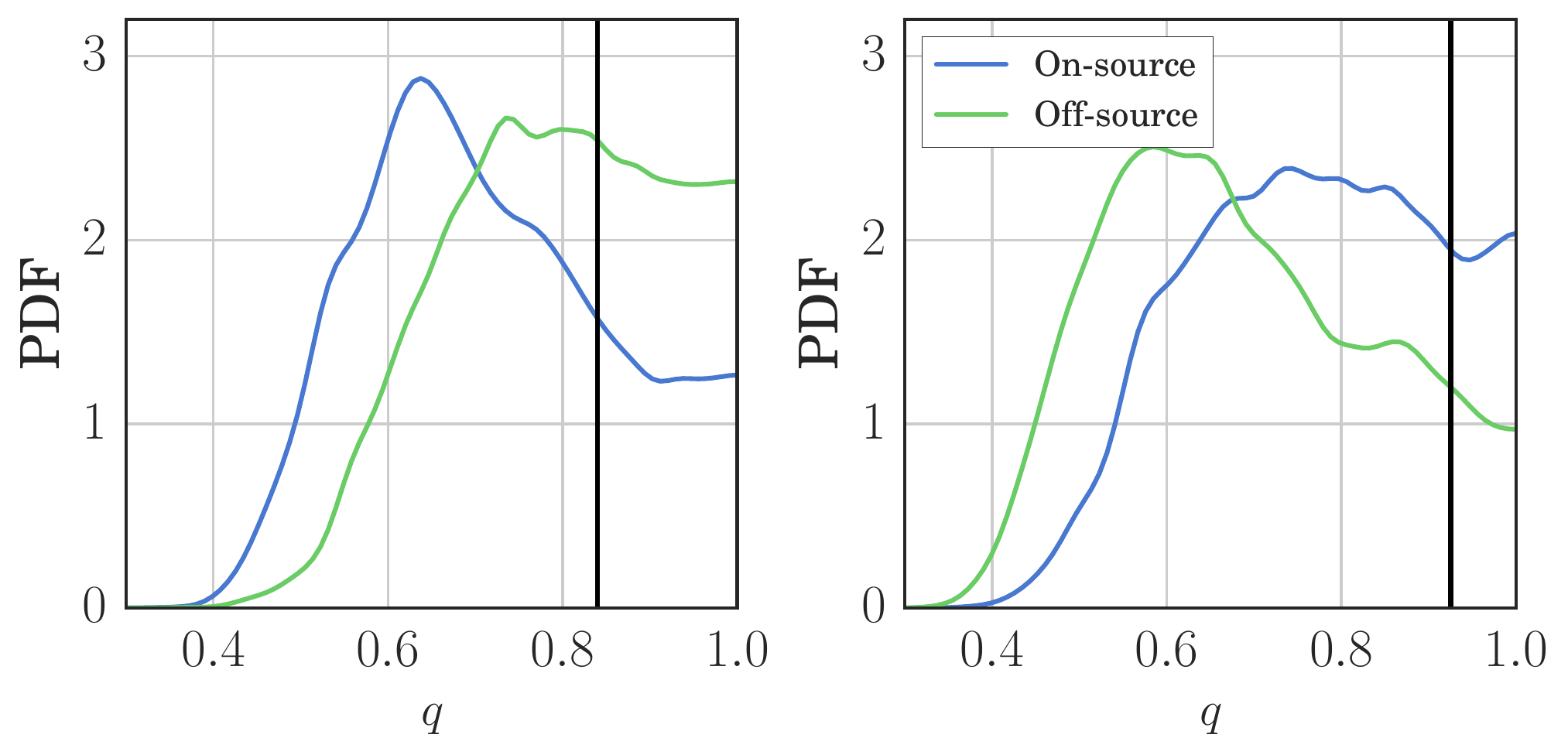}\\
\includegraphics[width=\columnwidth,clip=true]{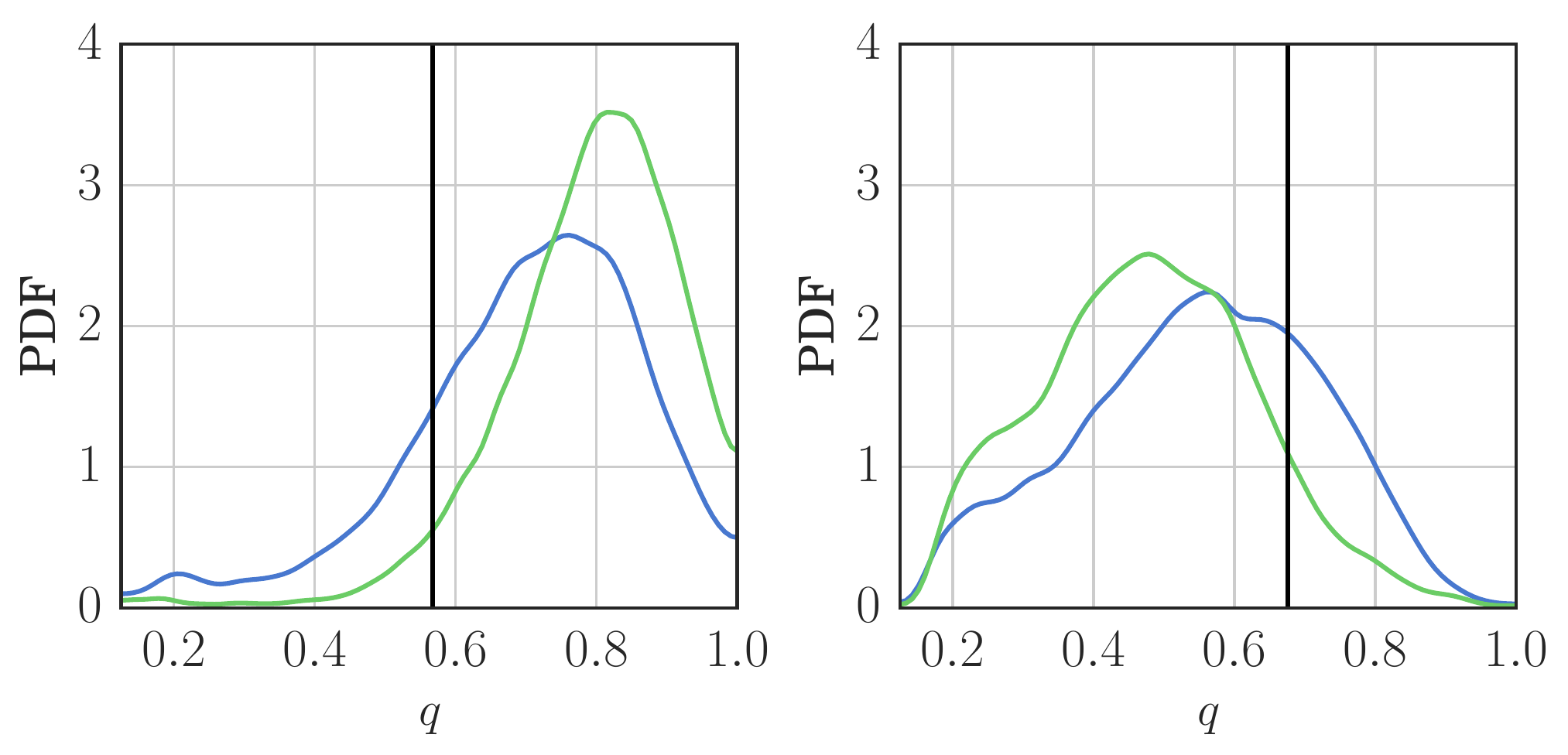}\\
\includegraphics[width=\columnwidth,clip=true]{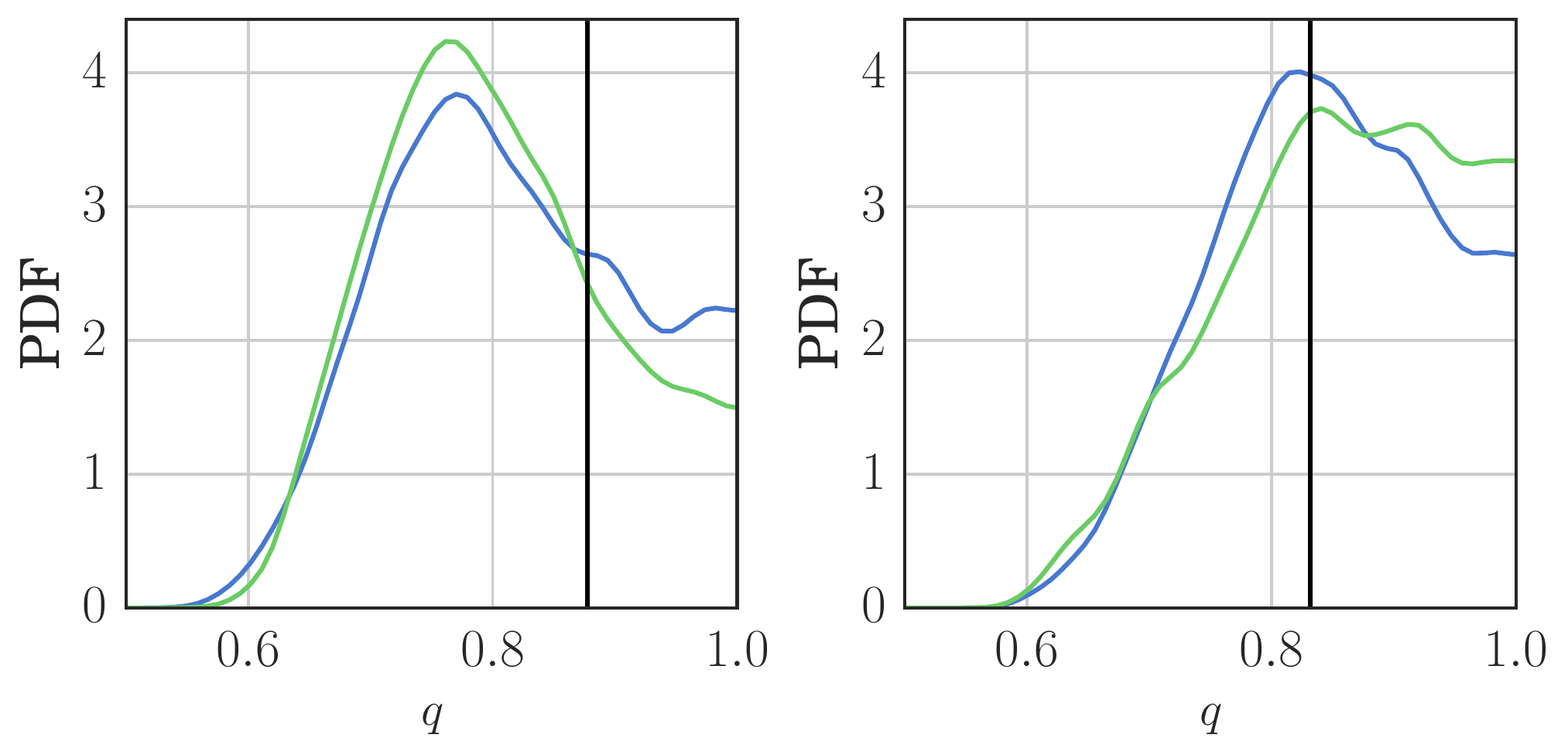}
\caption{Posterior distributions for the mass ratio obtained with the off-source and the on-source whitening filter for the two high mass BBH (top), low mass BBH (middle), and BNS (bottom) systems with the highest KS statistic from Fig~\ref{fig:KSAll}.
The true mass ratio is represented by the black vertical line.}
\label{fig:qAll}
\end{figure}

The mass ratio posteriors with the largest KS statistic are presented in Fig.~\ref{fig:qAll} where the vertical lines again denote the injected values.
The estimated mass ratio posteriors are fairly broad and in some cases extend to their prior bounds.
Nonetheless, we again find differences in the mass ratio posteriors that are comparable to other expected systematic errors, such as the omission of higher order modes~\cite{Chatziioannou:2019dsz}.

\begin{figure}[]
\includegraphics[width=\columnwidth,clip=true]{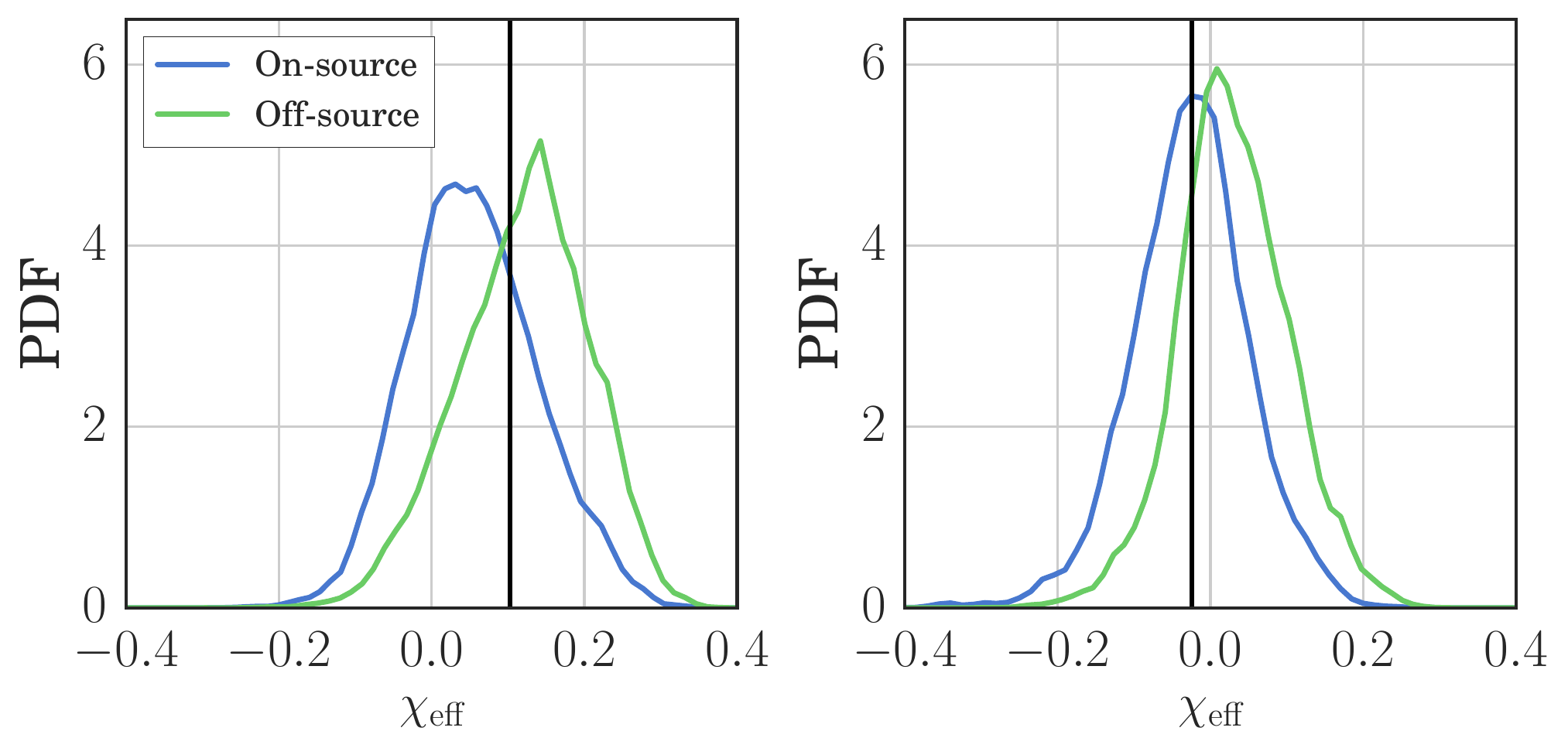}\\
\includegraphics[width=\columnwidth,clip=true]{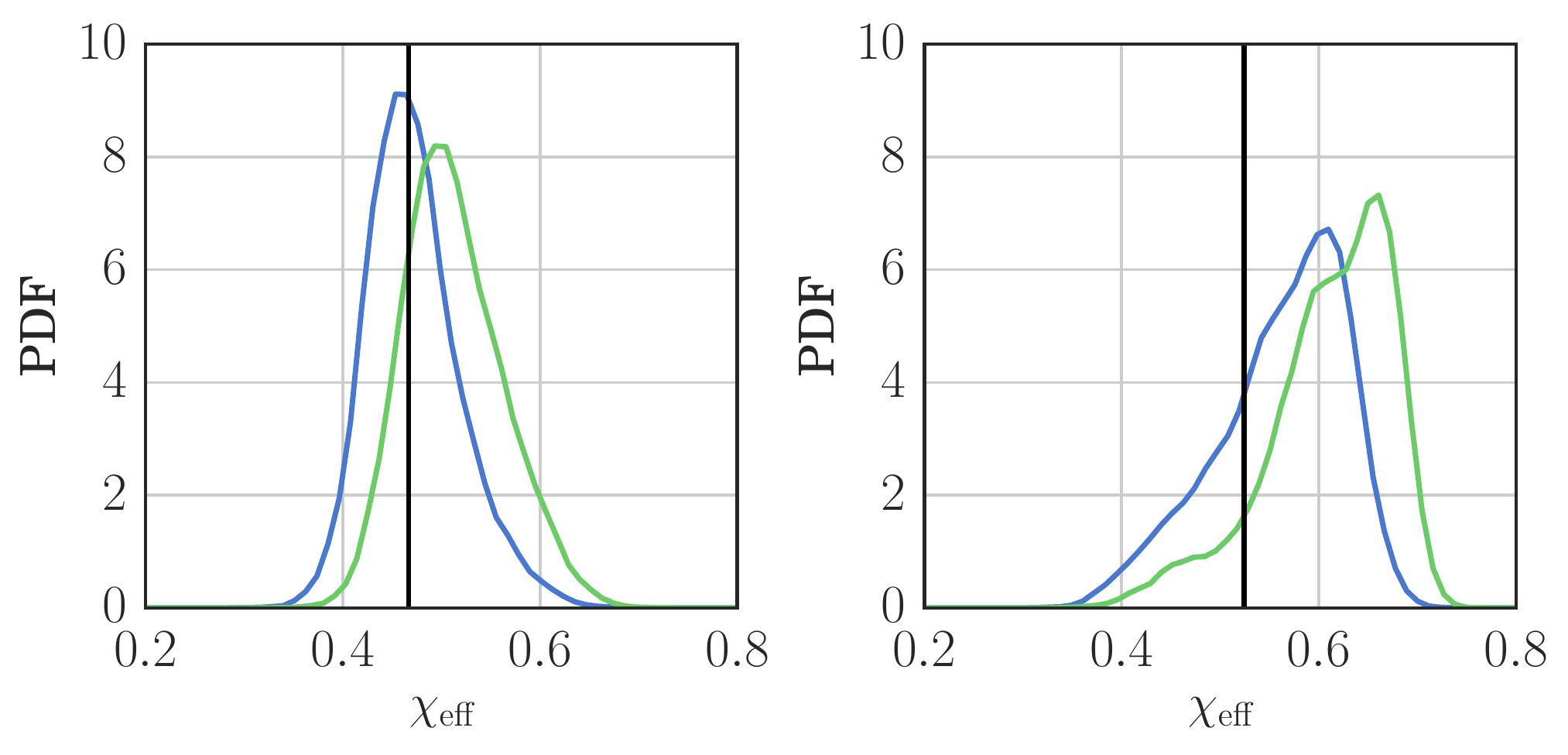}\\
\includegraphics[width=\columnwidth,clip=true]{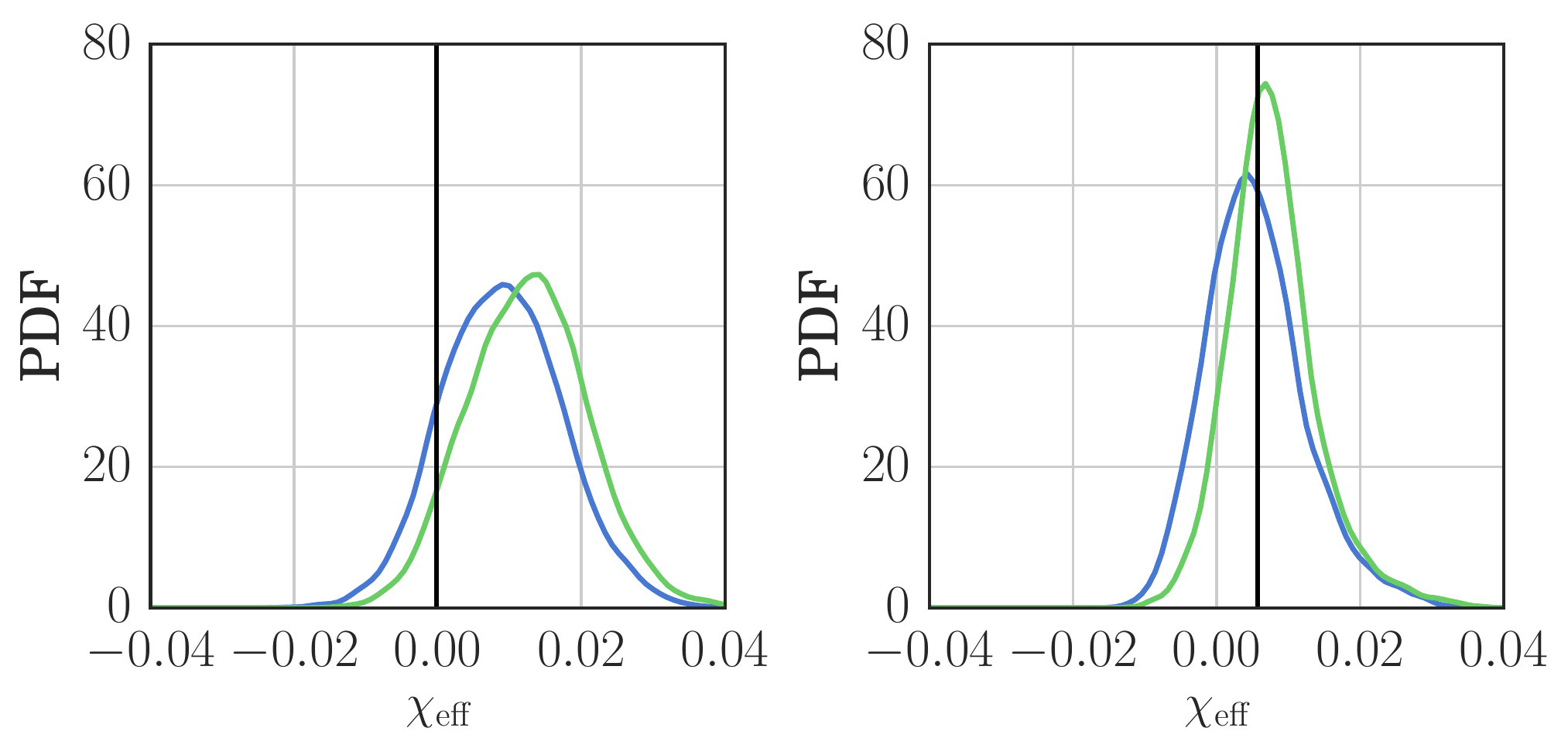}
\caption{Posterior distributions for the effective spin obtained with the off-source and the on-source whitening filter for the two high mass BBH (top), low mass BBH (middle), and BNS (bottom) systems with the highest KS statistic from Fig~\ref{fig:KSAll}.
The true effective spin is represented by the black vertical line.}
\label{fig:chieffAll}
\end{figure}

\begin{figure}[]
\includegraphics[width=.49\columnwidth,clip=true]{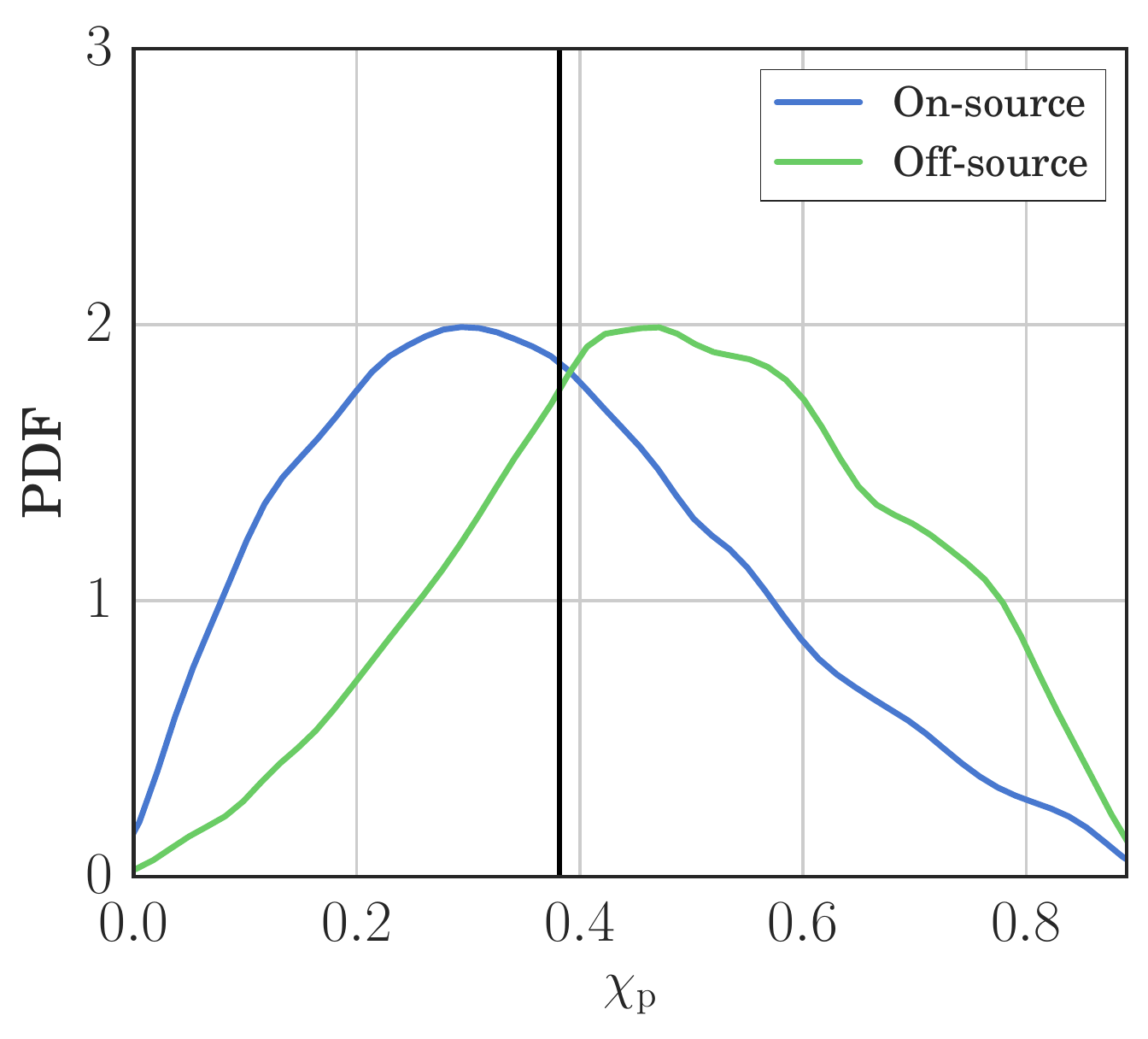}
\includegraphics[width=.49\columnwidth,clip=true]{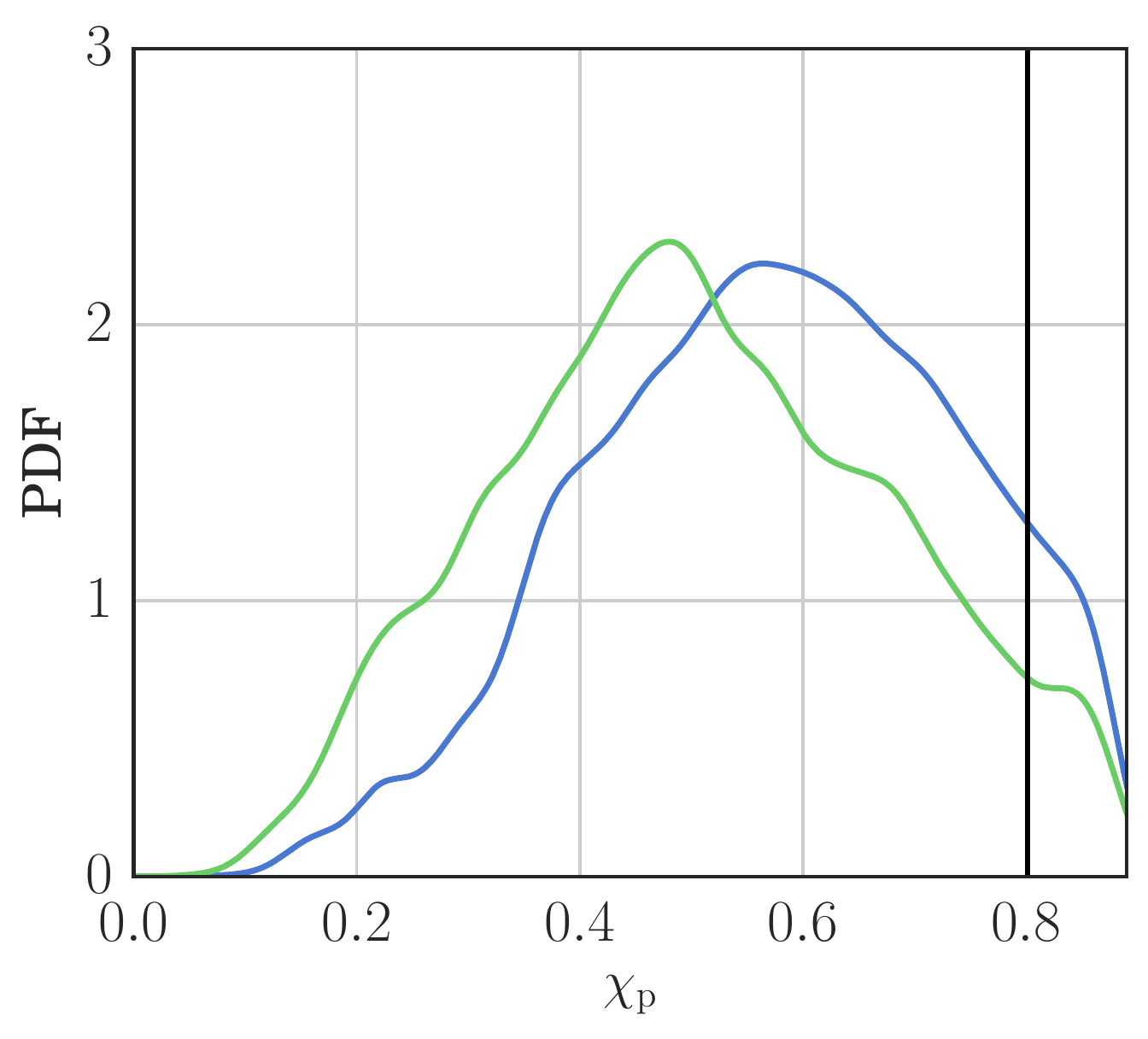}
\caption{Posterior distributions for the effective precession obtained with the off-source and the on-source whitening filter for the high mass BBH (left), low mass BBH (right) systems with the higher KS values from Fig~\ref{fig:KSAll}.
The true effective precession parameter is represented by the black vertical line.}
\label{fig:chipAll}
\end{figure}

The spin parameters are studied in Figs.~\ref{fig:chieffAll} and~\ref{fig:chipAll}. The effective spin posteriors, shown in Fig.~\ref{fig:chieffAll}, correspond to one of the best measured spin parameter combinations, though relatively poorly constrained compared to the masses.
We still find small shifts in the resulting posteriors, showing that the noise variance estimate can influence inference about the spin distribution of BHs, especially for loud or multiple events.
The effective spin-precession parameter $\chi_p$~\cite{Schmidt:2014iyl}, shown in Fig.~\ref{fig:chipAll}, is significantly less measurable than the effective spin.
In the worst case (left panel for a high mass BBH system) we find a large effect on the resulting posterior.

\begin{figure}[]
\includegraphics[width=\columnwidth,clip=true]{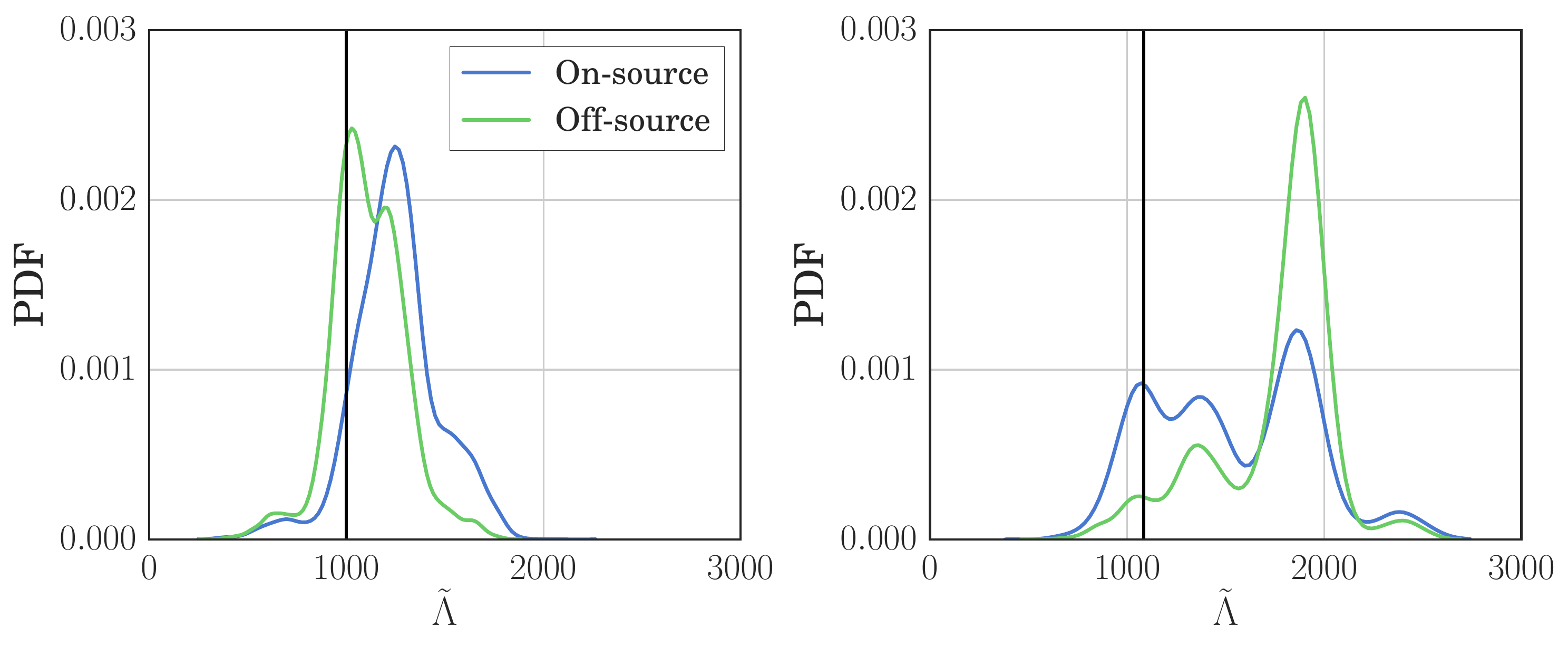}
\caption{Posteriors distributions for the effective tidal parameter obtained with the off-source and the on-source whitening filter for the BNS systems with the higher KS values from Fig~\ref{fig:KSAll}.
The true effective tidal parameter is represented by the black vertical line.}
\label{fig:LamtildeAll}
\end{figure}

Finally, the tidal parameters are in Fig.~\ref{fig:LamtildeAll} for the two BNS injections with the largest value of the KS statistic.
Even though the effective tidal parameter $\tilde{\Lambda}$ is not expected to be well measured (at least compared to the mass parameters) we again find examples where the posteriors are visibly different.
This might be due to the fact that $\tilde{\Lambda}$ is measured from a small time-frequency region, corresponding to the last few milliseconds of the emitted signal.
Therefore even small imperfections in the noise variance estimation in those frequencies could influence the $\tilde{\Lambda}$ posterior.

Besides the intrinsic parameters of the systems, we also notice that the off-source spectral estimation method leads to systematically larger values for the matched-filter \SNR of the signals.
The latter is defined as $(d|h)/\sqrt{(h|h)}$ and it is an estimate of both how well a template models the data and of the intrinsic loudness of the template. 
The off-source PSD estimation attempts to produce an unbiased estimate of $S(f)$ from the off-source data. 
The likelihood for $S(f)$ follows approximately an inverse-$\chi^2$ distribution; for such a distribution the mean is smaller than the median -- that is, the probability mass concentrates at ``small values''. 
An unbiased estimate of $S(f)$ will produce an estimate of $1/S(f)$ -- and therefore \SNR -- that is biased toward larger values. 
This in turn leads to generically larger values for the matched-filter \SNR. 
This consideration shows that \SNR is not an appropriate discriminator between different noise variance estimation methods.

\section{Stationarity}
\label{stationarity}

\begin{figure}[]
\includegraphics[width=\columnwidth,clip=true]{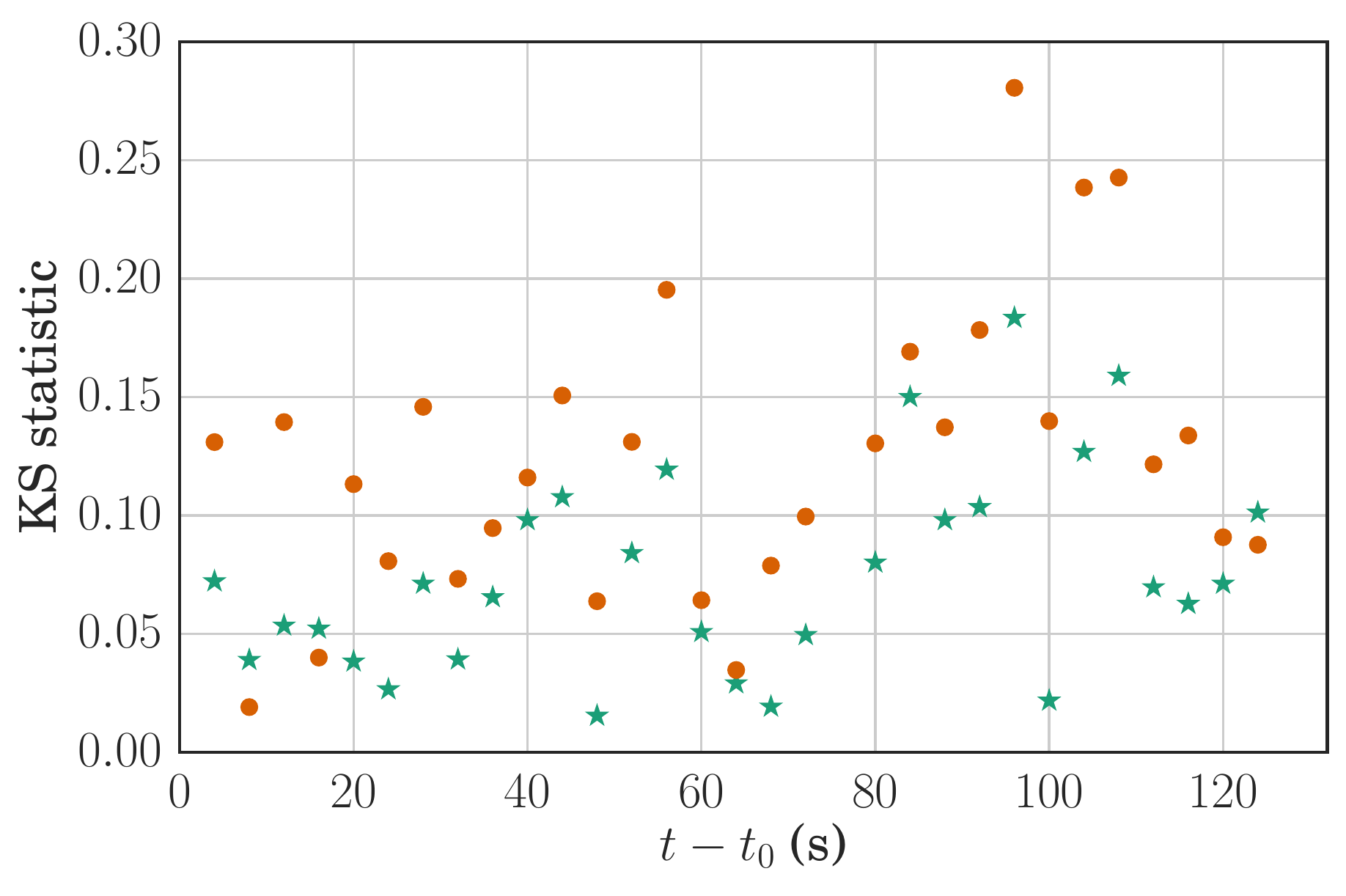}
\caption{KS statistic for the chirp mass of two high mass BBH events injected at $t_0$ and analyzed with whitening filters computed with the \BayesLine on-source methodology from data at $t-t_0$ as a function of the time delay $t-t_0$.
We find a generally increasing trend of the KS with the time distance between the signal and the data used to compute the whitening filter.}
\label{fig:stationarity}
\end{figure}

The whitening tests presented in Sec.~\ref{tests} suggest that the performance of the on-source filter degrades as the duration of the data segment analyzed increases.
Indeed for the BNS case (bottom row of Fig.~\ref{fig:ADAll}) the obtained on-source $A^2$ values are at the edge of the $3\sigma$ error.
To investigate this more we turn our attention to the assumption that the detector noise analyzed is stationary.

The stationarity assumption refers to the expectation that the mean and the variance of the noise does not change with time and it is instrumental in expressing the noise correlation matrix as a diagonal matrix, Eq.~\eqref{Lstationary}.
This assumption is expected to become less reliable with increasing data duration and it is known to completely break down for long stretches of data.
In fact, Littenberg and Cornish~\cite{Littenberg:2014oda} present evidence for deviations from stationarity on timescales of $\sim 64s$, already within the requirements of BNS PE analyses.

We here perform an exploratory study on the impact of nonstationarity on PE results. We select two of our simulated high mass BBH signals and inject them at time $t_0$ in the LIGO-Hanford detector.
We then use the \BayesLine algorithm to compute the on-source whitening filter from $4$s stretches of data that are increasingly removed from $t_0$.
In particular, we compute the whitening filter for data within $[t_0-2-4i,t_0+2-4i]$, for $i \in [0,32]$.
Case $i=0$ corresponds to the same analysis as Sec.~\ref{PE} and as performed in Ref.~\cite{LIGOScientific:2018mvr} for example.
Case $i=31$ corresponds to data about $128$s away from the signal, which is the duration of current BNS analyses.

 We then perform PE and compute the KS between the posterior for the chirp mass obtained from case $i=0$ and cases $i \in [1,32)$.
The result is plotted in Fig.~\ref{fig:stationarity} for both events and as a function of $4 i$, the time difference between the $i=0$ and each subsequent data segment.
The resulting KS values exhibit a general upward trend with time, suggesting that the effects of nonstationarity in the detector noise can have a noticeable impact on PE with segment durations as low as $128$s.
Given that such  segment durations are already in use and essential for analysis of BNS signals at current detector sensitivity, investigating ways to handle departures from noise stationarity is more pressing than usually assumed.

\section{Conclusions}
\label{conclusions}

This paper compares two methods for estimating the noise spectrum of Advanced LIGO data when analyzing short-duration transients:
A periodogram-based approach that uses data near in time to a candidate event versus a parametrized model that is fit to the data containing the candidate transient event.
The off-source method assumes the noise is Gaussian and stationary over the entire stretch of data used for spectral estimation and source characterization, while the ``on-source'' parametrized model makes the same assumptions but only over shorter stretches of data.

Comparisons of the noise estimation methods are designed in the context of parameter estimation applications, and are performed on data from Advanced LIGO's second observing run in which simulated signals have been added.
The simulated signals are representative of different merger events observed by LIGO-Virgo, including high mass BBH mergers (such as GW150914), low mass BBH mergers (such as GW151226) and BNS mergers (such as GW170817).

The Anderson-Darling statistic is employed to test assumptions about the noise implicitly encoded in the likelihood function used by LIGO-Virgo parameter estimation pipelines, particularly that it is stationary and Gaussian.
The Anderson-Darling tests provide indisputable evidence that the statistical properties of the data being analyzed are in better agreement with those assumed in the analysis methodology when using the parametrized model and on-source analysis (see Fig.~\ref{fig:ADAll}).
The conjectured cause of the difference in performance between methods is that the stationary timescales of the noise are shorter than the duration of data needed for the off-source spectral estimation.
This is supported by the result that the off-source method produces increasingly poorer fits to the noise as the duration of data needed for the analysis (and therefore needed for the spectral estimation) increases.

Note that the parametrized model also begins to diverge from the theoretical expectations for the Anderson-Darling tests when analyzing 128s of data (with the BNS simulations), suggesting that the assumptions about the noise properties are not supported over such durations.
This conclusion is also supported by the analysis of Fig.~\ref{fig:stationarity} which shows that parameter posteriors are increasingly affected as the distance in time between the signal and the data used for spectral estimation grows.
As ground-based GW detector sensitivities continue to improve, low-mass binaries will remain in the measurement band of the detectors for longer durations and analysis procedures need to adapt to these changes.

Having confirmed that the on-source method obeys better the Gaussianity assumption made by the likelihood function, 
we also study how differences in the spectral estimation map to inferences about the physical parameters of the source.
The Kolmogorov-Smirnov statistic is used to compare the inferred posterior distributions from data analyzed with the on-source and off-source noise estimates (see Fig.~\ref{fig:KSAll}).
From these comparisons it is clear that the choice of spectral estimation method does affect the inferred parameter distributions.
To see how these differences are manifested in the actual inferred parameters, Figs.~\ref{fig:McAll}-\ref{fig:LamtildeAll} show the marginalized posteriors for various parameters of particular interest corresponding to the results with the largest values of the KS statistic (i.e. the least similar).

Based on the parameter estimation results, we conclude that considering differences in spectral estimation methods is more than just an academic exercise.
It instead has measurable impact on inferences drawn from the data, and thus should be given the same scrutiny as other ingredients of the analysis, such as waveform models.

Given the sensitivity of parameter recovery to spectral estimation methods, it is clearly favorable to not only use a parametrized model and on-source estimation, but to also incorporate that model into the analysis and marginalize over its uncertainty, as is currently done in the \BayesWave\ pipeline for template-free detection and characterization of transients.
Such a capability is an extremely desirable feature to incorporate into the parameter estimation pipelines used for compact merger analyses.
In the absence of that capability, we recommend adopting methods such as those developed in \BayesLine\ for spectral estimation and, for the sake of reproducibility,  suggest the median noise spectrum as a suitable point estimate.

\acknowledgements

We thank Max Isi for comments on the manuscript.
This research has made use of data, software and/or web tools obtained from the Gravitational Wave Open Science Center (https://www.gw-openscience.org), a service of LIGO Laboratory, the LIGO Scientific Collaboration and the Virgo Collaboration.
LIGO is funded by the U.S. National Science Foundation.
Virgo is funded by the French Centre National de Recherche Scientifique (CNRS), the Italian Istituto Nazionale della Fisica Nucleare (INFN) and the Dutch Nikhef, with contributions by Polish and Hungarian institutes.
The authors are grateful for computational resources provided by the LIGO Laboratory and supported by National Science Foundation Grants PHY-0757058 and PHY-0823459, and for resources provided by the Open Science Grid~\cite{pordes:2007,Sfiligoi:2009}, which
is supported by the National Science Foundation award 1148698, and the U.S.
Department of Energy's Office of Science.
The Flatiron Institute is supported by the Simons Foundation.
C.-J. H. acknowledges support of the MIT physics department and the LIGO Laboratory which is operated under grant PHY-1764464 from the National Science Foundation.
S. G. acknowledges support from NSF grant PHY-1809572. 
J. A. Clark acknowledges support from NSFgrants OAC-1841479 and PHY-1700765.
N. C. acknowledges support from NSF grant PHY-1607343.
Parts of this research were conducted by the Australian Research Council Centre of Excellence for Gravitational Wave Discovery (OzGrav), through project number CE170100004.
Plots in this manuscript have been made with {\tt matplotlib}~\cite{Hunter:2007}.
%

\bibliography{OurRefs}

\end{document}